%% file: main.tex
\documentclass[fleqn,usenatbib]{mnras/mnras}

\usepackage{newtxtext,newtxmath}

\usepackage[T1]{fontenc}
\usepackage[normalem]{ulem}

\usepackage{graphicx}	
\usepackage{amsmath}	

\usepackage{amssymb}	

\input{newcommands.tex}

\title[Pair-instability black hole mass gap]{
    Formation of GW190521 from stellar evolution: the impact of the hydrogen-rich envelope, dredge-up and \CagO\ rate on the pair-instability black hole mass gap   
}
  
\author[Costa et al.]{
Guglielmo Costa$^{1,2,3}$ \thanks{E-mail: guglielmo.costa@unipd.it},
Alessandro Bressan$^{4,3}$,
Michela Mapelli$^{1,2,3}$,
Paola Marigo$^{1}$,
Giuliano Iorio$^{1,2,3}$, \newauthor
Mario Spera$^{4}$
\\
$^{1}$Dipartimento di Fisica e Astronomia Galileo Galilei, Universit\`a di  Padova, Vicolo dell'Osservatorio 3, I--35122 Padova, Italy\\
$^{2}$INFN-Padova, Via Marzolo 8, I--35131 Padova, Italy\\
$^{3}$INAF-Padova, Vicolo dell'Osservatorio 5, I--35122 Padova, Italy\\
$^{4}$SISSA, via Bonomea 365, I--34136 Trieste, Italy
}
    
\date{Accepted XXX. Received YYY; in original form ZZZ}

\pubyear{2020}

\begin{document}
\label{firstpage}
\pagerange{\pageref{firstpage}--\pageref{lastpage}}
\maketitle

\begin{abstract}
Pair-instability (PI) is expected to open a gap in the mass spectrum of black holes (BHs) between $\approx{}40-65$~M$_\odot$ and $\approx{}120$~M$_\odot$. The existence of the mass gap is currently being challenged by the detection of GW190521, with a primary component mass of $85^{+21}_{-14}$~M$_{\odot}$. Here, we investigate the main uncertainties on the PI mass gap: the \CagO\ reaction rate and the H-rich envelope collapse. With the standard \CagO\ rate, the lower edge of the mass gap can be 70~M$_\odot$ if we allow for the collapse of the residual H-rich envelope at metallicity $Z\leq{}0.0003$. 
Adopting the uncertainties given by the {\sc starlib} database, for models computed with the \CagO\ rate $-1\, \sigma$, we find that the PI mass gap ranges between $\approx{}80$~M$_\odot$ and $\approx{}150$~M$_\odot$. Stars with $M_{\rm ZAMS}>110$ M$_\odot$ may experience a deep dredge-up episode during the core helium-burning phase, that extracts matter from the core enriching the envelope. As a consequence of the He-core mass reduction, a star with $M_{\rm ZAMS} =160$~\Msun\ may avoid the PI and produce a BH of 150~\Msun. In the $-2\,{}\sigma{}$ case, the PI mass gap ranges from 92~M$_\odot$ to 110~M$_\odot$. Finally, in models computed with \CagO\ $-3\,{}\sigma{}$, the mass gap is completely removed by the dredge-up effect. The onset of this dredge-up is particularly sensitive to the assumed model for convection and mixing. The combined effect of H-rich envelope collapse and low \CagO\ rate can lead to the formation of BHs with masses consistent with the primary component of GW190521.
\end{abstract}

\begin{keywords}
            convection --
            stars: evolution --
            stars: mass loss --
            stars: massive --
            stars: interiors --
            stars: black holes
\end{keywords}



%
%

\section{Introduction}
\label{sec:Intro}
    The LIGO-Virgo collaboration (LVC) recently reported the discovery of GW190521 \citep{abbottGW190521,abbottGW190521astro}. With a primary black hole (BH) mass of $85^{+21}_{-14}$ M$_{\odot}$ and a secondary BH mass of $66^{+17}_{-18}$ M$_{\odot}$ (90\% credible intervals), this binary black hole (BBH) merger is the most massive one observed with gravitational waves to date \citep{abbottGW150914,abbottO1astro,abbottO1,abbottO2,abbottO2popandrate,abbottGW170817,abbottGW190412,abbottGW190425,abbottGW190814}. The mass of the primary BH is a puzzle for astrophysicists, because it lies in the middle of the pair-instability (PI) mass gap. 
    
    PI is possibly the key process to understand the maximum mass of stellar-origin BHs, because it is expected to carve a gap in the mass spectrum of BHs between $\approx{}40-65$ M$_\odot$ and $\approx 120$ M$_\odot$ (e.g., \citealt{Heger2002,woosley2007,Belczynski2016pair,Woosley2017,Woosley2019,Spera2017,Spera2019,Giacobbo2018,mapelli2019,tanikawa2020a}). The uncertainty on the boundaries of the PI mass gap depends on a number of physical processes (e.g., \citealt{Takahashi2018,Marchant2019,Stevenson2019,Leung2019,Farmer2019,Farmer2020,Mapelli2020,Renzo2020,Vanson2020,tanikawa2020b}).
     
    PI starts to be efficient when the plasma inside stars reaches temperatures above $6 \times 10^8$ K and densities between about $10^2$ and $10^6$ g cm$^{-3}$. Such conditions on internal temperature and density are reached when massive CO cores are formed in stars. Pair creation converts thermal energy into rest mass of $e^-\,e^+$, lowering the central pressure. This causes an hydro-dynamical instability  that leads to core collapse. In the collapsing core, oxygen is ignited explosively. 
   
    In stars developing a helium core mass $\approx{}32-64$~M$_\odot$ at the end of carbon burning, PI manifests as pulsational PI (PPI): the star starts to pulsate losing a large amount of mass, until it finds a new equilibrium. After PPI, the star will end its life with a core-collapse supernova or a direct collapse, leaving a BH. More massive stars, with helium core mass $\approx{}64-135$~M$_\odot$, undergo a PI supernova (PISN): a powerful single pulse that totally destroys the star, leaving no compact remnant. Finally, PI triggers the direct collapse of stars with He core mass $>135$ M$_\odot$, eventually producing an intermediate-mass BH \citep{Heger2002}. 
    Such theoretical picture has been confirmed by many authors in both 1-D \citep{Heger2002, Woosley2017, Takahashi2018, Marchant2019} and multi-dimensional hydrodynamical simulations \citep{Chatzopoulos2013, Chen2014}.
    From an observational perspective, iPTF14hls \citep{Arcavi2017} and SN 2016iet \citep{Gomez2019} cannot be explained by current core-collapse supernova models and might be connected with PI.
    
    If a massive star enters or not in the PI regime depends mainly on the CO core mass left after the central He burning and on its evolution during the following advanced phases. For a given zero-age main sequence (ZAMS) mass (M$_{\rm ZAMS}$) and stellar metallicity ($Z$), the final CO mass can be strongly affected by many evolutionary processes, such as mass-loss, convection, overshooting, semi-convection \citep{Kaiser2020, Clarkson2020, Renzo2020}, rotation \citep{Chatzopoulos2012,limongi2018,Song2020,Mapelli2020,marchant2020} and internal and surface magnetic fields \citep{Haemmerle2019, Keszthelyi2020}.
    These processes may change the ZAMS mass versus CO  core mass relation and, thus, the initial mass at which stars enter in the PI regime \citep{Takahashi2018}.
    Another important ingredient that influences both the CO core mass and its composition is represented by nuclear reactions during the core helium burning (CHeB). The two most important ones are the triple-$\alpha$ reaction, $^4$He($2 \alpha$, $\gamma$)$^{12}$C, and the carbon-$\alpha$ one, \CagO. 
    The interplay of these two reactions during the CHeB phase determines the final carbon-to-oxygen ratio (\COf) in the CO core, hence the final fate of the star \citep{Weaver1993}.
    The \CagO\ reaction rate is one of the most uncertain \citep{Caughlan1988, Buchmann1996, Sallaska2013, deBoer2017,Rapagnani2017}, because it is difficult to measure.
    The study on how the \CagO\ reaction affects the stellar evolution is an old problem \citep{Iben67, Brunish90, Alongi91}. In recent years, many efforts have been done to understand its importance on the most advanced stellar phases and in the final fates \citep[][]{Woosley2007b,Tur2007, Tur2010, deBoer2017, Fields2018, Takahashi2018, Farmer2019, Sukhbold2020, Farmer2020}.

    In particular, \cite{Farmer2019} and \cite{Farmer2020} show that the \CagO\ reaction rate is one of the main sources of uncertainty on the boundaries of the PI mass gap, when pure helium stars are considered. Another major source of uncertainty on the final BH mass is the fate of the residual hydrogen envelope in case of a direct collapse: if a massive star retains a fraction of its H envelope to the very end of its life, is this envelope able to collapse together with the internal layers of the star? Or is it so loosely bound that it is expelled, even without a supernova explosion \citep{Sukhbold2016}? If the residual H envelope collapses to the final BH, the lower edge of the PI mass gap ($M_{\rm gap}$) might increase by $\approx{}20$ M$_\odot$ \citep{Mapelli2020}.

    The main focus of this work is to study the effect of the uncertainty of the \CagO\ reaction rate on the $M_{\rm gap}$ in metal-poor stars. With respect to \cite{Farmer2020}, who consider only pure-helium stars, we also investigate stars with hydrogen envelopes. This enables us to model the PI mass gap accounting for the possible collapse of the residual H envelope. Moreover, we find that the onset of a dredge-up during the H shell burning phase plays a crucial role for the evolution of the CO core of the most massive stars. The PI mass gap might even disappear for the combined effect of envelope undershooting and low \CagO\ reaction rate. 
    Our goal is to determine under which conditions the formation of BHs like the primary component of GW190521 \citep{abbottGW190521,abbottGW190521astro} is possible through single stellar evolution.
    
    In Section~\ref{sec:implemented_physics} we give a description of the {\sc parsec} code, we describe the method adopted to include different \CagO\ reaction rates and how we test the stellar dynamical stability during the evolution. 
    In Section~\ref{sec:Results_He}, we present the new {\sc parsec} evolutionary tracks of pure-He stars computed with varying \CagO\ reaction rates and compare them with the results by other authors.
    In Section~\ref{sec:Results_H}, we present the new tracks of massive stars with hydrogen envelopes and compare the maximum masses found for the $M_{\rm gap}$ with other studies. Finally, in section~\ref{sec:Conc} we draw our conclusions.

%
%

\section{Methods}
\label{sec:implemented_physics}
    In this work we use the {\sc parsec} V2.0 code that has been extensively described in \citet{Costa2019a} and references therein. In the following, we describe the major updates adopted for this work.

        \subsection{Winds of massive stars}
        \label{sec:mloss}
            Mass-loss by stellar winds is modelled as described in \cite{Chen2015}. For massive hot stars, we take into account the dependence of mass loss on both stellar metallicity \citep{Vink2000,Vink2001} and Eddington ratio \citep{Graefener2008,Vink2011}. 
            With this mass loss the models are able to reproduce the Humphreys–Davidson limit \citep{Humphreys1984} observed in the Galactic and Large Magellanic Cloud (LMC) colour–magnitude diagrams\footnote{We do not have calibrations at lower metallicity than the LMC one, thus, an uncertainty on the mass loss remains. A recent study by \citet{Jiang2018} on mass outburst on luminous blue variable stars using detailed 3-D simulations found outburst episodes that lead to instantaneous mass-loss rate of $\approx{}0.05$ \Msun yr$^{-1}$, at solar metallicity.} \citep{Chen2015}.
            The metallicity dependence of mass loss is expressed as a dependence on the surface iron abundance\footnote{Thus, when using the fitting formulas from  \citet{Vink2011},  the iron content must be re-scaled to the iron content assumed for the Sun in \citet{Vink2001}.}.

            For Wolf-Rayet (WR) stars, we adopt the new revised mass-loss prescription by \citet{Sander2019}:
            \begin{equation}\label{eq:sander}
                {\rm Log} \frac{\dot{M}}{{\rm M}_\odot\,{} {\rm yr}^{-1}} = -8.31 + 0.68 \, {\rm Log} \frac{L}{L_\odot}.
            \end{equation}
            
            This relation is a good fit for Galactic WR type-C (WC) and WR type-O (WO) stars (as shown in  fig. 5 of \citealt{Sander2019}) but, as such, it does not include a metallicity dependence which is of paramount importance for our work. In order to include a metallicity dependence in Eq.~\ref{eq:sander}, we use the results by \cite{Vink2015}. From models of WN and WC stars at varying Fe and C--O abundances, \cite{Vink2015} concludes that the main driver of mass-loss is Fe and that the eventual surface C excess in WC stars only produces a lower limit threshold for the mass-loss rate. By fitting \cite{Vink2015} models of WN and WC stars at varying surface Fe abundance, we derive two multiplicative factors (for WN and WC stars respectively) to be applied to  Eq.~\ref{eq:sander}: 
            \begin{eqnarray}
                f_\text{WN}&=& - 1 + 1.9 \tanh \left\{0.58 \left[{\rm Log}(X_{\rm Fe}) + 1 \right]\right\}, \\
                f_\text{WCO}&=& - 0.3 + 1.2 \tanh \left\{0.5 \left[{\rm Log}(X_{\rm Fe}) + 0.5 \right]\right\}
            \end{eqnarray}
            where $X_{\rm Fe}$ is the mass fraction iron content. 
            
            These two factors are obtained by fitting models of WN and WC/WO stars in which mass-loss mainly depends on iron and, to a lesser extent, on CNO elements \citep{Vink2015}. When metallicity decreases, the mass-loss rate lowers and flattens to values given by the CNO elements \citep[see fig. 2 in][]{Vink2015}. 
            It is unlikely that the WR mass-loss depends on the helium surface content which, instead, is likely the result of a high mass-loss rate \citep{Bestenlehner2014}.

        \subsection{Opacity, neutrinos and equation of state}
        \label{sec:EOS}
            In the high-temperature regime, 4.2 $\leq$ log (T/K) $\leq$ 8.7, we adopt the opacity tables provided by the Opacity Project At Livermore (OPAL)\footnote{http://opalopacity.llnl.gov/} team \citep[][and references  therein]{Iglesias1996}, in the low-temperature regime, 3.2 $\leq$ Log (T/K) $\leq$ 4.1, we employ the \textsc{\ae sopus} tool\footnote{http://stev.oapd.inaf.it/aesopus} \citep{Marigo2009}. Conductive opacities are included following \citet{Itoh2008}.
            
            Energy losses by electron neutrinos are taken from \citet{Munakata1985} and \citet{Itoh1983}, and for plasma neutrinos we use the fitting formulae by \citet{Haft1994}.

            To compute the equation of state (EOS) of stellar matter, we adopt two different models, depending on central temperature.
            For temperature Log (T/K) < 8.5,  we adopt the \textsc{freeeos}\footnote{\url{http://freeeos.sourceforge.net/}} code version 2.2.1 by Alan W. Irwin. As already described in \cite{Bressan2012}, this code is integrated within our evolutionary code so we can obtain the EOS in two ways: we can compute it on-the-fly, with a higher degree of accuracy, but with a higher computational cost; or we can retrieve it by means of pre-computed look-up tables with different combinations of abundances. As shown in \citet{Bressan2012}, this second method is accurate enough and we prefer it for numerical speed reasons. At a given partition of elements ($X_{\rm i}/Z$) and metallicity, the tables are divided in two sets: i) the H-rich set in which there are 10 tables with a varying hydrogen and helium abundances, and ii) the H-free one composed of 12 tables with a varying helium, carbon and oxygen composition.
            
            For temperature Log (T/K) > 8.5, we adopt the  code described in \cite{Timmes1999}, because the {\sc freeeos} tables do not include the treatment of pair creation.

        \subsection{Nuclear reaction network}
        \label{sec:nuc_net}
            The nuclear reaction network consists in 72 different reactions and 33 isotopic elements, from hydrogen to zinc. The reaction rates and Q-values are taken from the {\sc jina} reaclib database \citep{Cyburt2010}. In the network we treat all the most important reactions from hydrogen to oxygen burning. The network includes also the reverse reactions of the $\alpha$-capture. 
            The new adopted reactions with their rates are listed in Table~\ref{tab:reacrate}. The other reactions, already included in the previous versions of the code, can be found in Table 1 of \citet{Fu2018}. 
            In the network, we include all the reactions of the cold-CNO cycle, but not those of the hot-CNO cycle. The latter requires high temperatures ($>$ 0.1 Gk) in H-rich regions to ignite, but these are never reached in our models. 
            
            In {\sc parsec} v2.0, nuclear reaction network and element mixing are treated and solved at the same time, adopting an implicit diffusive scheme \citep[more details in][]{Marigo2013, Costa2019a}.

            \begin{table} 
                    \caption{List of the nuclear reaction rates added to {\sc parsec} v2.0 in this work.
                    } 
                    \begin{center}
                    \begin{tabular}{lc} 
                        \hline\hline
                        Reaction & Reference \\
                        \hline
                        $^{16}$O($^{16}$O, $\alpha$)$^{28}$Si  & \citet{Caughlan1988}    \\    
                        $^{28}$Si($\alpha$, $\gamma$)$^{32}$S  & \citet{Cyburt2010}    \\   
                        $^{32}$S($\alpha$, $\gamma$) $^{36}$Ar & \citet{Cyburt2010}    \\   
                        $^{36}$Ar($\alpha$, $\gamma$)$^{40}$Ca & \citet{Cyburt2010}    \\    
                        $^{40}$Ca($\alpha$, $\gamma$)$^{44}$Ti & \citet{Cyburt2012}    \\    
                        $^{44}$Ti($\alpha$, $\gamma$)$^{48}$Cr & \citet{Cyburt2010}    \\    
                        $^{48}$Cr($\alpha$, $\gamma$)$^{52}$Fe & \citet{Cyburt2010}    \\    
                        $^{52}$Fe($\alpha$, $\gamma$)$^{56}$Ni & \citet{Cyburt2010}    \\    
                        $^{56}$Ni($\alpha$, $\gamma$)$^{60}$Zn & \citet{Cyburt2010}    \\    
                        \hline 
                    \end{tabular}
           			\end{center}
                     \footnotesize{This table reports only the reactions that were added to {\sc parsec} in this work. The other reactions, already included in the previous versions of the code, can be found in Table 1 of \citet{Fu2018}.
                     }
                \label{tab:reacrate} 
            \end{table}

    \subsection {The \CagO\ reaction}
        \begin{table} 
            \caption{
            Multiplier factors, \fco, for the \CagO\ reaction rates adopted in this work. See text for details.} 
            \centering 
            \begin{tabular}{cccccccc} 
                \hline\hline
                &$+3\, \sigma$ & $+2\, \sigma$ & $+1\, \sigma$ & $0\,\sigma$  & $-1\, \sigma$ & $-2\, \sigma$ &  $-3\, \sigma$ \\
                \hline
                \fco & 2.74 & 1.96 & 1.4 & 1.0 & 0.71 & 0.51 & 0.36 \\
                \hline 
            \end{tabular} 
            \label{tab:SigTable} 
        \end{table}
        For the \CagO\ reaction we use the  
        chw0 (v5) rate from the {\sc jina} reaclib database \citep{Fu2018}, which is a re-evaluation of the \citet{Buchmann1996} rate that includes high-lying resonances \citep{Cyburt2012}\footnote{Relative differences with respect to the most recent {\sc jina} reference rate ({\sc nacre}, v9) in the temperature range 0.1 -- 0.5 GK are about 3\%.}.

        The {\sc jina} database does not include uncertainties on the chw0 (v5) rate. To estimate the uncertainties, we compared our adopted rate with the one given in the {\sc starlib} database \citep{Sallaska2013}. In the {\sc starlib} database, each rate is given as a mean value with a correspondent confidence error at $1\, \sigma$. The \CagO\ rate given in the {\sc starlib} database is based on the rate by \citet{Kunz2002}.
        
\begin{figure}
            \includegraphics[width=0.48\textwidth]{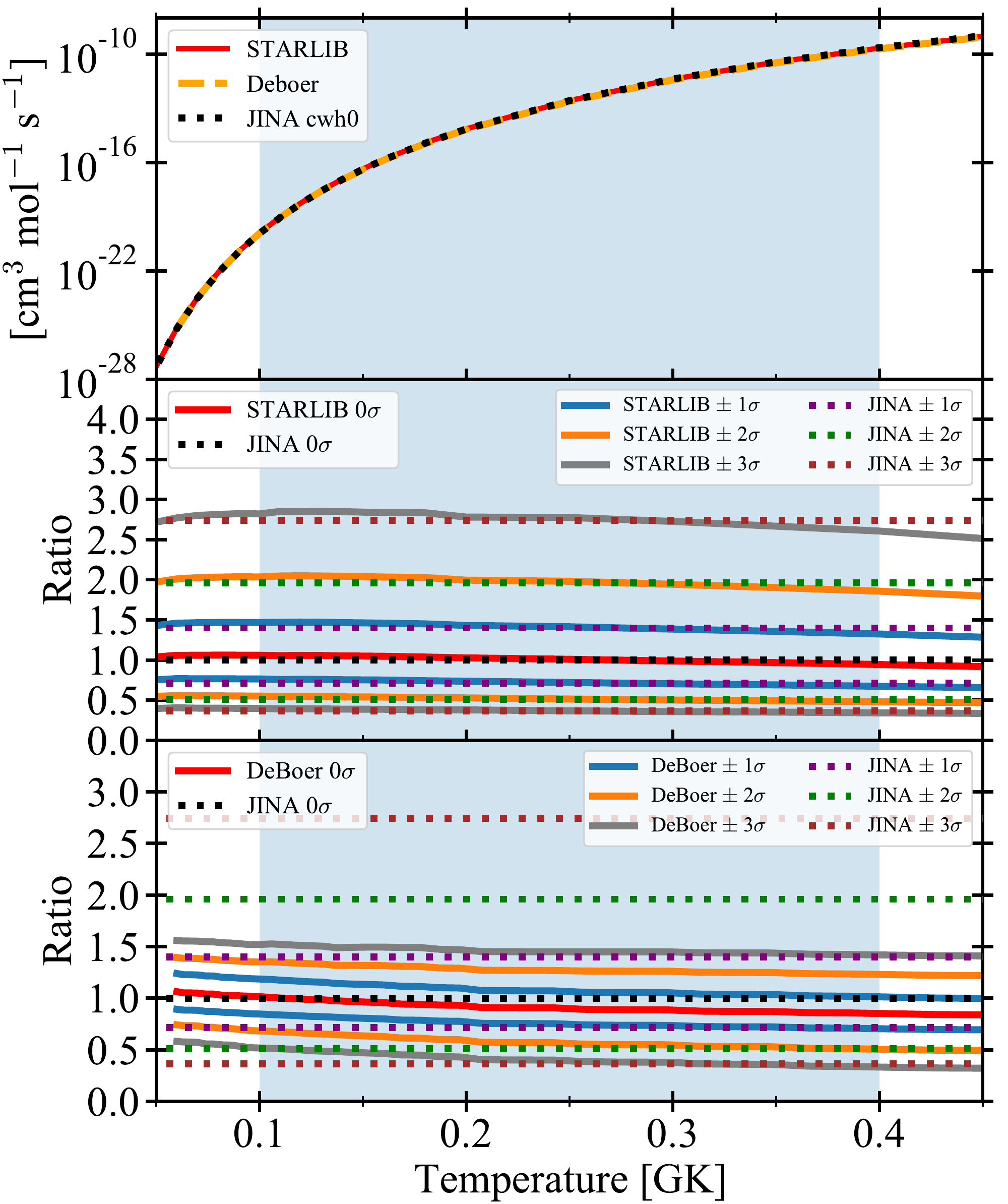}
            \caption{
                Upper panel: comparison of the \CagO\ rate from {\sc starlib} database, \citet{deBoer2017} and {\sc jina} database in red, orange and black, respectively. The blue shaded area indicates the temperature range of helium burning in the stellar core. 
                Middle panel: comparison between {\sc starlib} and {\sc jina} reaction rates with different $\sigma$. All the rates are divided by the standard ($0\, \sigma$) {\sc jina} rate. 
                The blue, orange and grey continuous lines are the {\sc starlib} rates with $\pm 1\, \sigma$, $\pm 2\, \sigma$ and $\pm 3\, \sigma$, respectively.
                The purple, green and brown dotted lines indicate the {\sc jina} $\pm 1 \,\sigma$, $\pm 2\, \sigma$ and $\pm 3\, \sigma$ rates, respectively. 
                Lower panel: comparison between \citet{deBoer2017} and {\sc jina} reaction rates with different $\sigma$.
                    }
            \label{fig:rates}
        \end{figure}

        Fig.~\ref{fig:rates} shows the comparison of the two rates in the temperature range of the core helium burning, i.e. between about 0.1 and 0.4 GK. The mean values of the two rates are in agreement within a few percent. To reproduce the uncertainties given by the {\sc starlib} \CagO\ reaction rate in the range between $-3\, \sigma$ and $+ 3\, \sigma$ we multiplied our adopted {\sc jina} rate by a temperature independent multiplier factor, \fco. 
        The values of \fco\ are listed in Table~\ref{tab:SigTable}. The middle panel of Fig.~\ref{fig:rates} shows the ratios of the rates with respect to the {\sc jina} standard and at different $\sigma$.
        In the bottom panel of Fig.~\ref{fig:rates}, we also compared the adopted uncertainties with those given by \citet{deBoer2017}. We find that the \citeauthor{deBoer2017} rates are generally lower than our adopted rates in the case of $+3\,\sigma$, $+2\,\sigma$ and $+1\,\sigma$. In the best value case, the two rates are very similar at low temperature but start to be different at high temperatures. Interestingly, the \citeauthor{deBoer2017} rates with $-1\,{}\sigma$, $-2\, \sigma$ and $-3\, \sigma$ show a very similar trend to the {\sc starlib} ones, at temperatures $>$ 0.2 Gk. Hence, had we used the uncertainties from \cite{deBoer2017}, our main conclusions on the maximum possible mass of a BH  (Sec.~\ref{sec:MassSpec}) would not have changed significantly.
        
    \subsection{Pair creation and dynamical instability}
    \label{sec:dyn_inst}
        The electron-positron creation process may induce dynamical instability during the most advanced phases of massive star evolution. Pair creation absorbs part of the thermal energy of the plasma, consequently lowering the thermal pressure.
        The plasma is not a perfect gas anymore, and temperature variations do not lead to changes in pressure \citep{Kippenhahn2012}. Regions in which this process happens become locally dynamically unstable.
        To check if a star is globally stable or not, a perturbation method should be adopted \citep{Ledoux1958}. However, \citet{Stothers1999} showed that an evaluation of the first adiabatic exponent properly weighted and integrated over the whole star, \G1{}, is a very good approximation to determine the dynamical stability of a star. As done by other authors \citep[e.g.][]{Marchant2019, Farmer2019, Farmer2020}, we adopted the \citet{Stothers1999} stability criterion, which states that a star is stable if
        \begin{equation}
            \G1{} = \frac{\int^{M}_0 \frac{\Gamma_1 P}{\rho} dm}
                         {\int^{M}_0 \frac{P}{\rho} dm} > \frac{4}{3},
        \label{eq:G1}
        \end{equation} 
        where $\Gamma_1$ is the first adiabatic exponent, $P$ is the pressure, $\rho$ is the density and $dm$ is the element of mass. We decided to compute the above integrals in two ways. In the first case, the integral is calculated from the centre of the star ($M = 0$) up to the mass of helium core ($M=M_{\rm He}$), defining \G1{Core}; in the second case, we compute the integral from the centre to the surface of the star ($M = M_{\rm TOT}$), to include the contribution of the envelope, thus \G1{TOT}. These two values are computed at each time-step for all tracks. In the case of pure-He stars, \G1{Core} = \G1{TOT}. 
        Since {\sc parsec} is a hydro-static code, we cannot follow the evolution through the dynamical collapse; we stop the computation if the \G1{} $< 4/3 + 0.01$, to be conservative, and label the star as a PI. Since we cannot follow the hydrodynamical evolution, we cannot distinguish between PPI and PISN: we classify both of them as PI and we assume that both of them leave no compact object. This makes our results for the BH mass even more conservative, because stars that undergo a PPI might still retain most of their mass after weak pulses and form a massive BH by core collapse \citep{Marchant2019, Farmer2019}.
        
        To improve the readability of \G1{} listed in Tables, we define the following value $\G1{-} \equiv{} \G1{} - (4/3 + 0.01)$. When $\G1{-} > 0$, we consider the star globally stable, otherwise, the star is unstable. Following the evolution through hydro-dynamical phases is beyond the purpose of this paper. 
    %
    
%
%
\section{Models of pure-He stars}
\label{sec:Results_He}
    
    We first analyze pure-He stars and compare our results with other studies.
    \subsection{Evolutionary tracks}
    \label{sec:Tracks_He}
    %
            %
        \begin{figure}
            \includegraphics[width=0.48\textwidth]{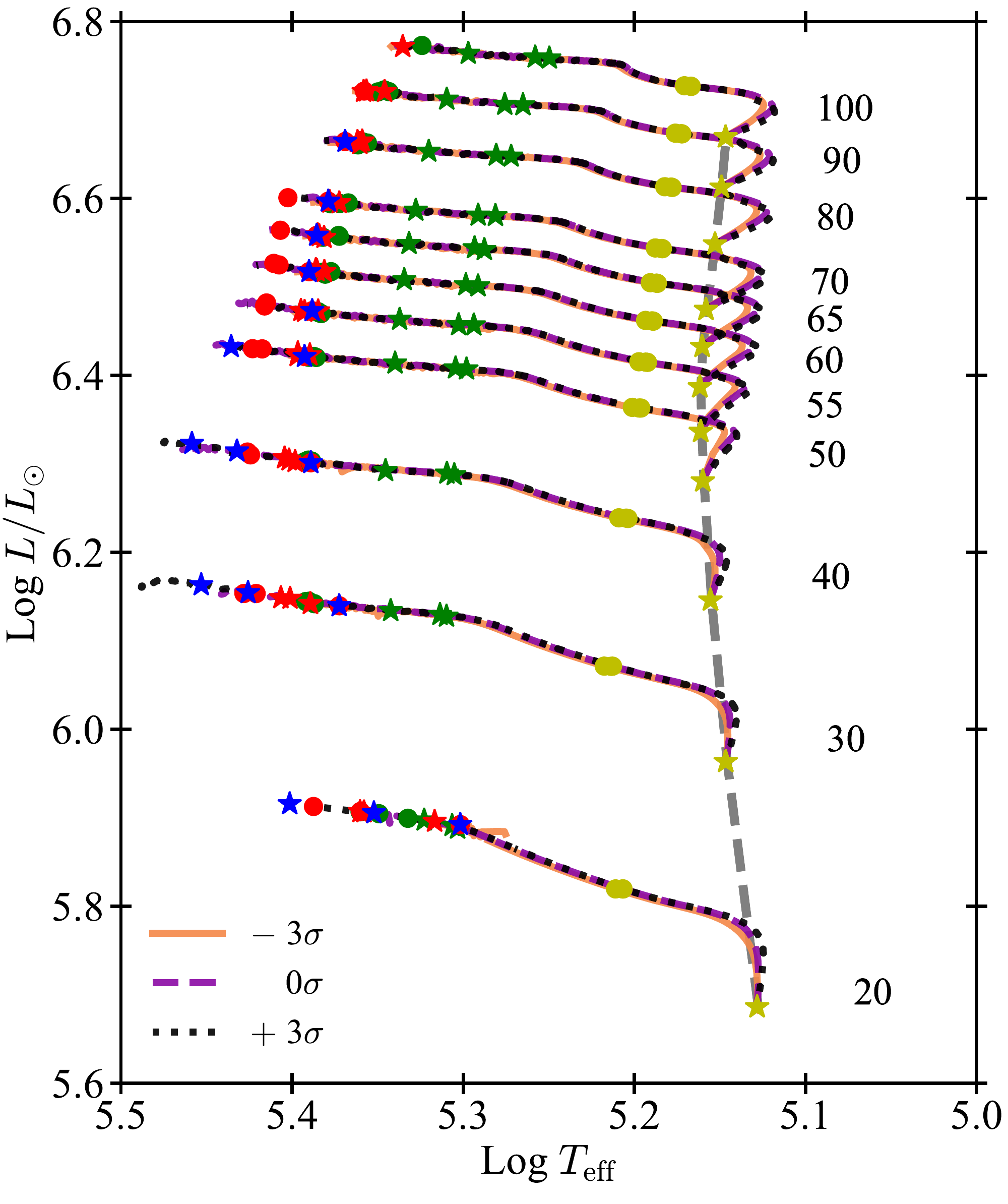}
            \caption{
                HR diagram of three sets of tracks of pure-He stars, computed with Z = 0.001. Solid black lines, dashed green lines and dotted brown lines indicate tracks computed with the \CagO\ reaction rate $-3\,\sigma$, with $0 \, \sigma$ and with $+3\,\sigma$, respectively. The dashed gray line shows the pure He-ZAMS. The symbols indicate helium (yellow), carbon (green), neon (red) and oxygen (blue) burning phases, respectively. Stars and circles symbols indicate the ignition and depletion of each element, respectively. The numbers on the right of the ZAMS line indicate the initial mass in~\Msun\ units.
                    }
            \label{fig:PureHe_HR}
        \end{figure}
        %
                %
        \begin{figure}
            \includegraphics[width=0.48\textwidth]{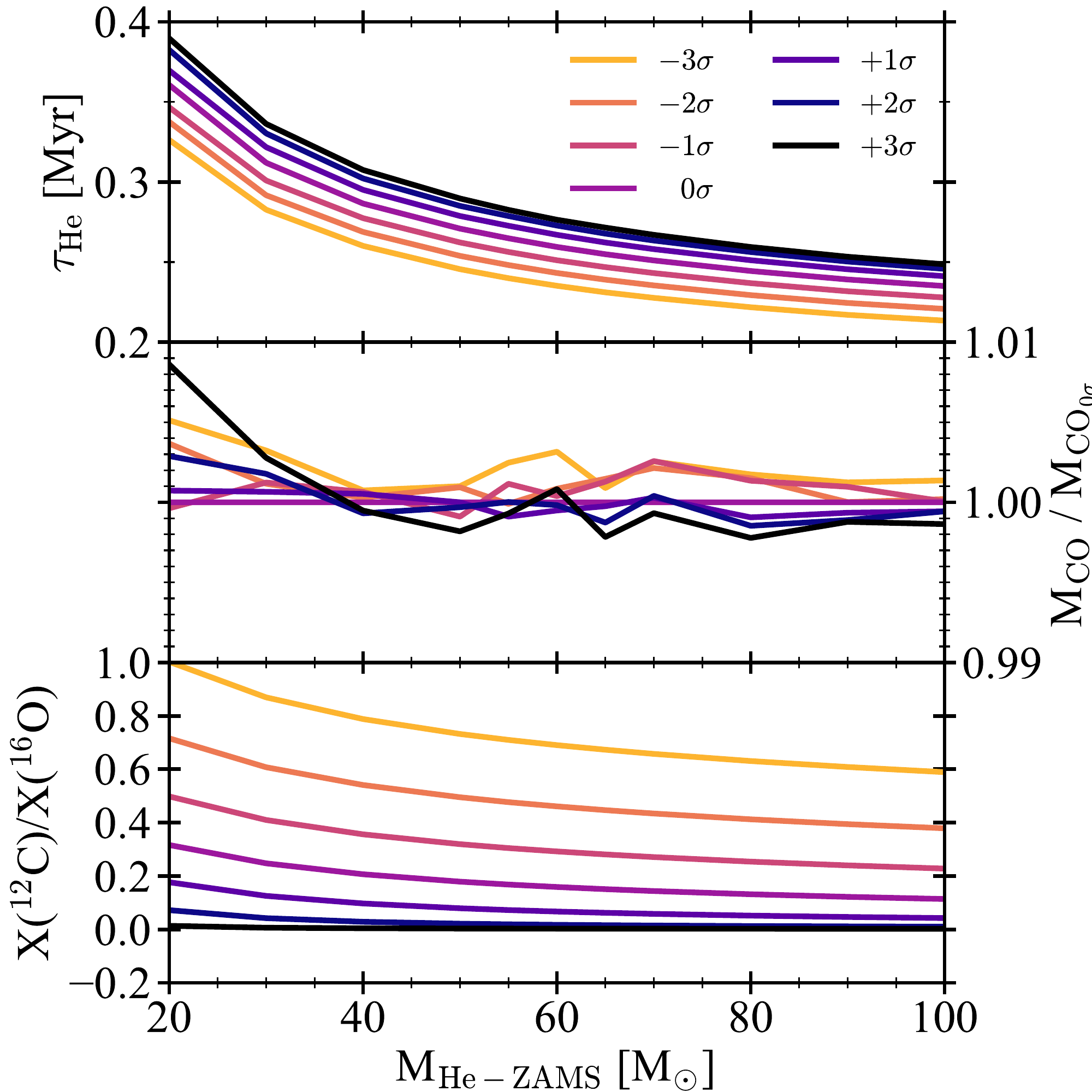}
            \caption{Upper panel: He-MS lifetimes versus the initial mass of pure-He stars for different adopted rates. Middle panel: Ratio between the CO core masses and the CO core mass of the model computed with $0 \, \sigma$ at the end of the He-MS. Lower panel: carbon-to-oxygen ratio in the CO cores at the end of the He-MS. 
                }
            \label{fig:PureHe_ts}
        \end{figure}
        %
        %
        \begin{figure*}
            \includegraphics[width=0.33\textwidth]{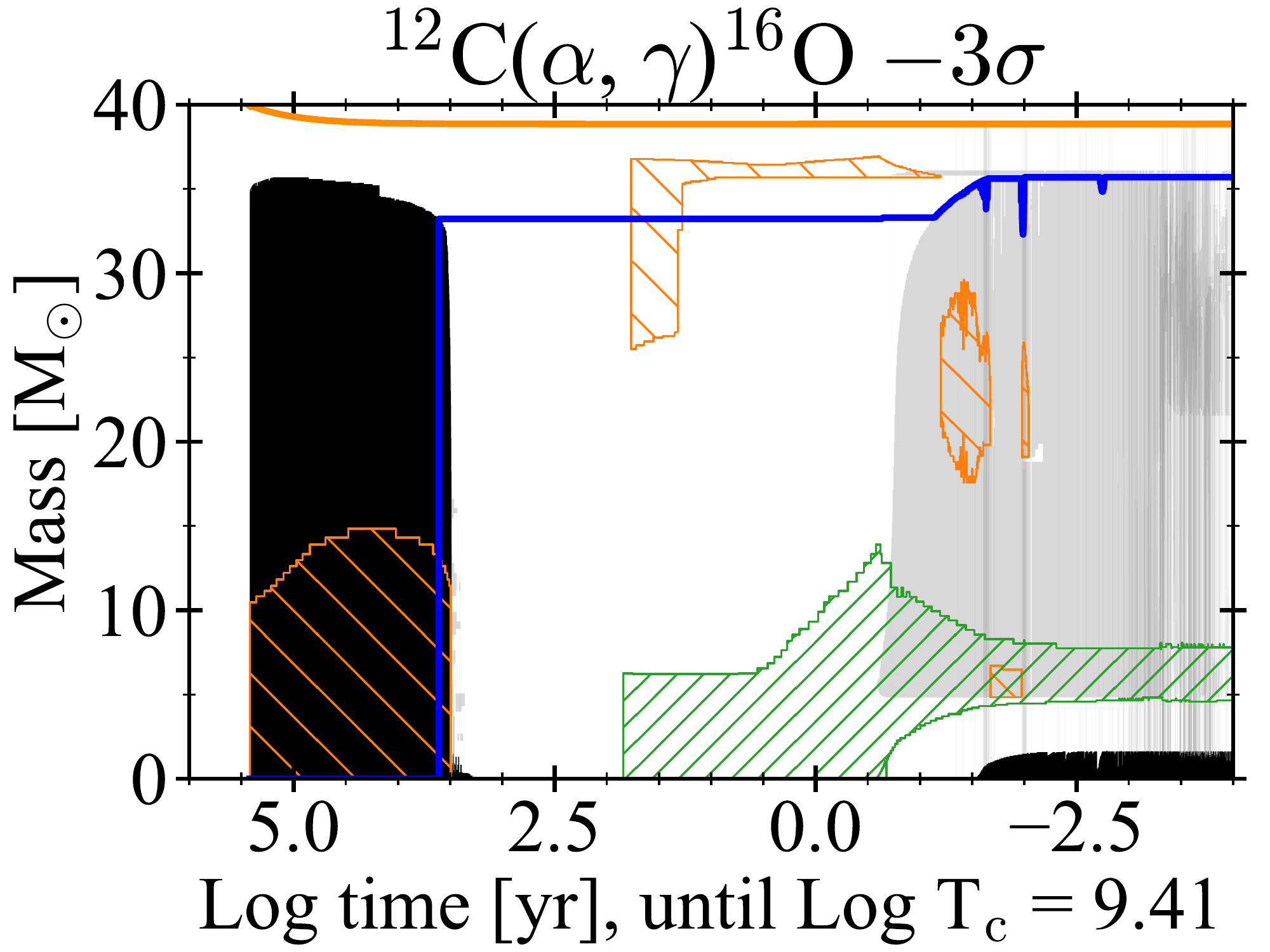}
            \includegraphics[width=0.33\textwidth]{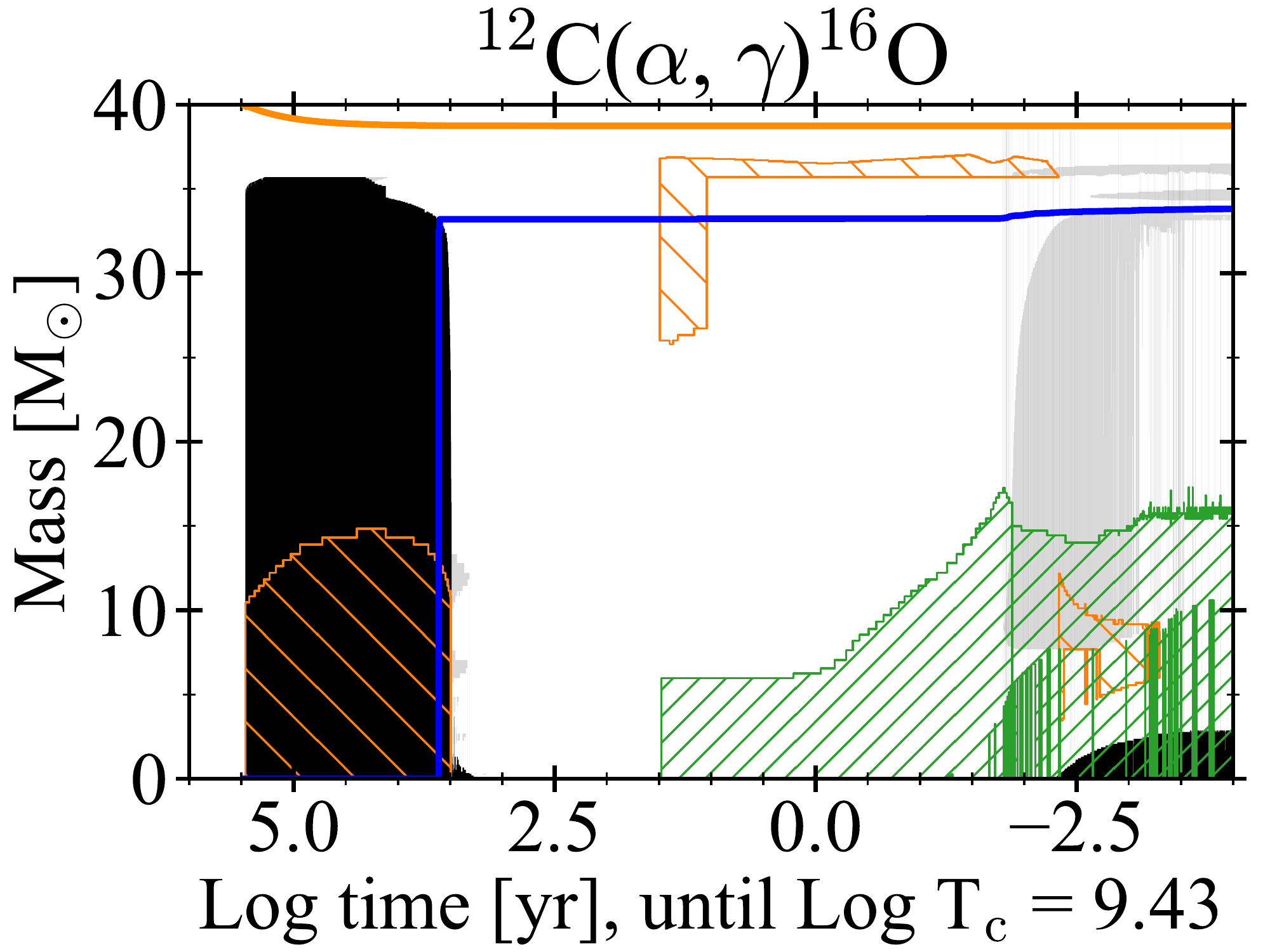}
            \includegraphics[width=0.33\textwidth]{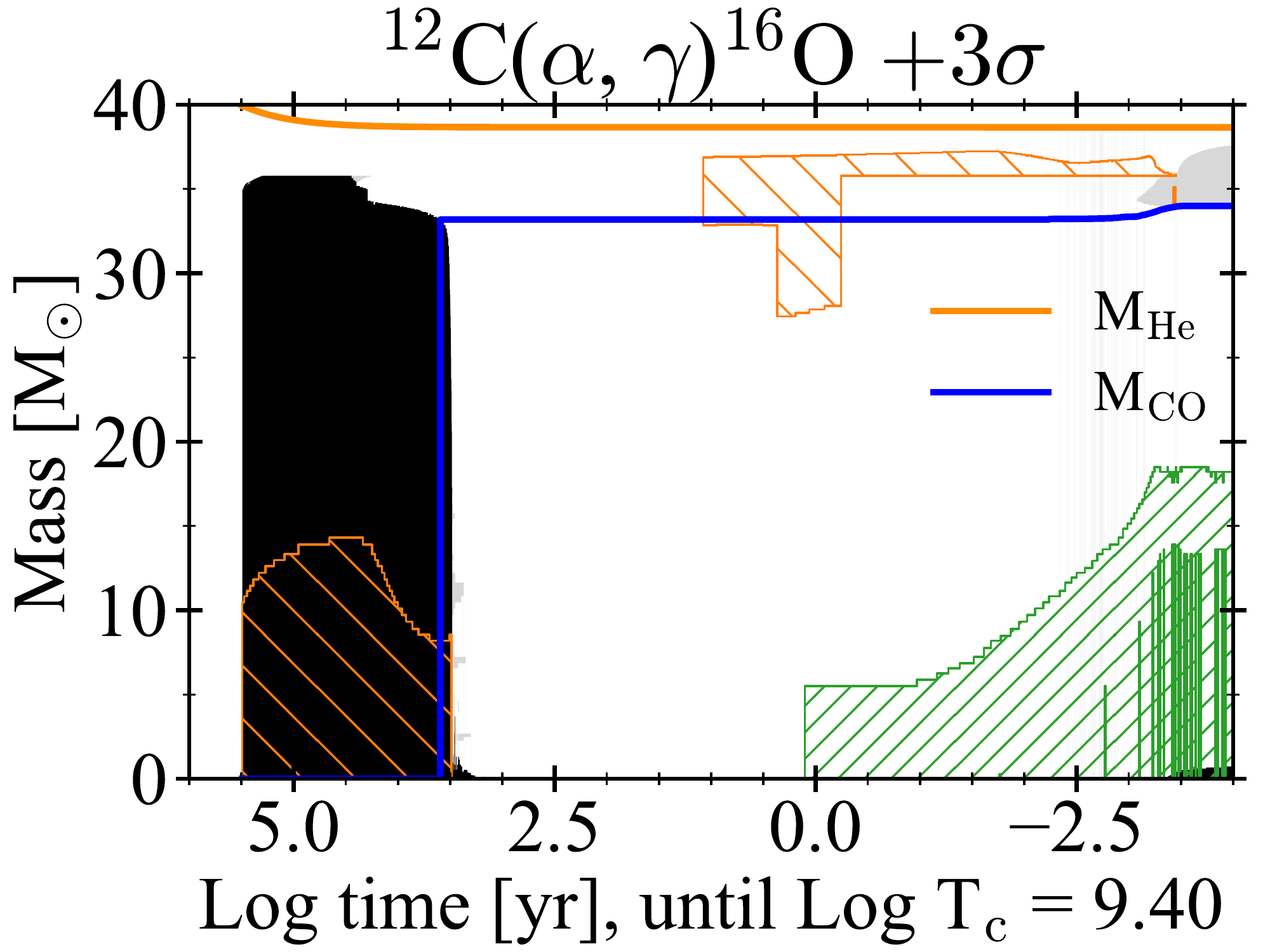}

            \caption{Kippenhahn diagrams of three models of pure-He stars with M$_{\rm ZAMS}=40$\Msun\ and with the \CagO\ reaction rate $- 3 \,\sigma$, $0 \, \sigma$ and with $+ 3\, \sigma$ in the left-hand, middle and right-hand panels, respectively. The black areas represent the convective unstable core, while the grey areas indicate the convective envelope or intermediate convective regions between the core and the surface of the star.
            Continuous yellow and blue lines show the stellar mass and the CO core. 
            The yellow and green hatched areas indicate the helium and carbon burning regions, respectively. We plot zones that contribute at least for the 1\% of the L$_\mathrm{He}$ and L$_\mathrm{C}$ luminosity at a given time-step.
             }
            \label{fig:PureHe_Kipp}
        \end{figure*}
        %
        %
        \begin{figure}
            \includegraphics[width=0.48\textwidth]{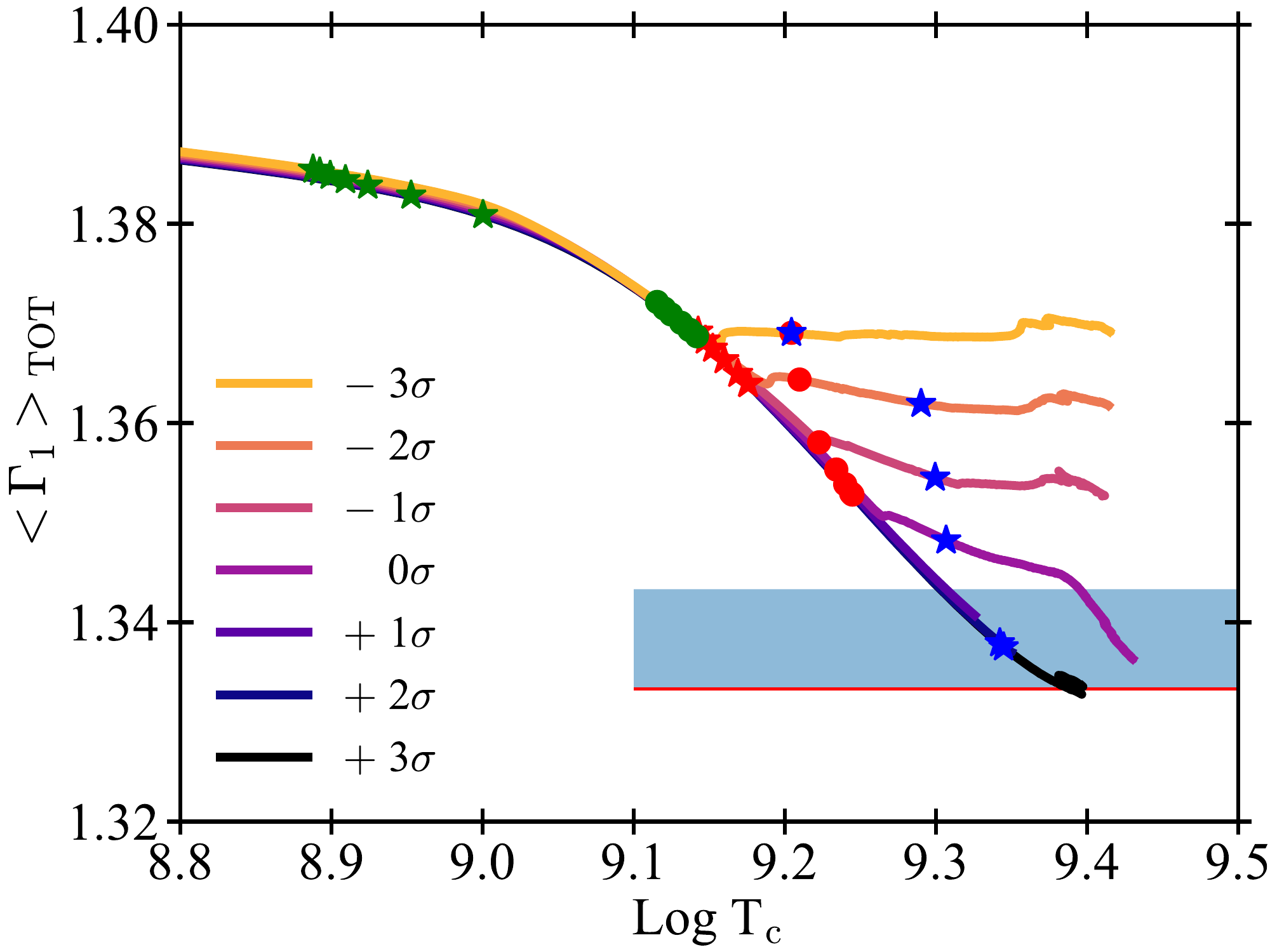}
            \caption{Comparison of the averaged first adiabatic exponent, \G1\, versus the central temperature of pure-He stars models with M$_{\rm ZAMS}$ = 40~\Msun. The plot shows the final phases at the end of the evolution. Coloured lines indicate models computed with different \CagO\ reaction rates. Symbols indicate the start and the end of core burning phases of elements with the same colour code as adopted in Fig.~\ref{fig:PureHe_HR}. The red horizontal line corresponds to $\G1{}=4/3$. The above blue shaded area indicates values between 4/3  and 4/3 + 0.01, range in which the whole star starts to be dynamically unstable.
                    }
            \label{fig:PureHe_gamma1}
        \end{figure}
    %
        We computed pure-helium evolutionary tracks with seven different values of the multiplier parameter, \fco, for the \CagO\ reaction rate, that correspond to varying the rate between $-3\, \sigma$ and $+3\, \sigma$ \citep{Sallaska2013}. Rotation is neglected in this study, since it increases the CO  core and enhances the mass-loss during the evolution, with the effect to lower the final masses \citep{Chatzopoulos2012, Mapelli2020,Song2020}.
        All tracks evolve from the helium ZAMS (He-ZAMS) to  core oxygen burning, when stars remain globally stable (\G1{TOT-} > 0). 
        
        To build the He-ZAMS we adopted the same methodology as described by \citet{Spera2019}. The new sets are composed of stars with the following masses 20, 30, 40, 50, 55, 60, 65, 70, 80, 90 and 100~\Msun. We adopted the solar-scaled element mixture by \citet{Caffau2011}, and a metallicity of Z~=~0.001. We choose this particular value of metallicity to better compare our results with those found by \citet{Farmer2019, Farmer2020}.
        For the mixing treatment, we adopt the mixing length theory \citep[MLT,][]{Bohm-Vitense1958} with a fixed solar calibrated value of the mixing length, \amlt\ = 1.74 \citep{Bressan2012}. We adopt the Schwarzschild criterion to define the convective regions, and a core overshooting with a mean free path of the convective element across the border of the unstable region of \lov~=~0.4 in units of pressure scale height, \Hp. Therefore, we are assuming an overshooting distance of $\approx 0.2$~\Hp\ above the convective core \citep{Costa2019a}.
       
        In the convective envelope we adopt an undershooting distance of $\Lambda_{\rm env}$~=~0.7~\Hp\ below the border of the unstable region.
        This value of $\Lambda_{\rm env}$ is calibrated on observations. In particular, it allows us to reproduce the location of the red giant branch (RGB) bumps and their luminosity function, and the extension of blue loops of intermediate-mass stars, in colour-magnitude diagrams of globular clusters \citep{Alongi91, Tang2014, Fu2018}.
        \citet{Tang2014} showed that higher values of $\Lambda_{\rm env}$ might be needed in order to better reproduce the observed blue-to-red supergiants ratios in low metallicity dwarf irregular galaxies, while \citep{Fu2018} suggested that an even higher value should be used to fit the RGB bumps of the most metal poor globular clusters.

        Fig.~\ref{fig:PureHe_HR} shows the evolution of three selected sets computed with \fco\ corresponding to $+ 3\, \sigma$, $0\, \sigma$ and $-3 \,\sigma$.
        Stars begin their evolution from the He-ZAMS. We define such phase as the moment in which the star is fully sustained by the nuclear reactions' energy, and the star has burnt less than 1\% of the helium hosted in the core. During the helium main sequence (He-MS) all stars evolve to higher luminosity. 
        Stars with M$_{\rm ZAMS} < 50$~\Msun\ evolve directly toward hotter effective temperatures. Stars with M$_{\rm ZAMS} \geq 50$~\Msun\ first expand reducing their effective temperature and then turn to higher temperatures. As the evolution proceeds, the stars move to higher luminosity and hotter effective temperature and start to burn and deplete sequentially carbon, neon and then oxygen in their cores.
        
        Fig.~\ref{fig:PureHe_HR} shows that different rates do not affect the path followed by stars in the HR-diagram, but instead they may affect when and where stars ignite and deplete carbon, neon and oxygen in their cores. These small variations are due to different mass fractions of carbon and oxygen present in the cores at the end of the He-MS, that depend on the assumed rate of the \CagO\ reaction.  
        The evolution is eventually stopped if the star becomes globally unstable by checking the \G1{TOT-} value defined in Eq.~\ref{eq:G1}. 
        In the other case, we continue the evolution until the lowest possible value of central oxygen is reached. These values are shown in Table \ref{tab:HeTable}. At this point, we stop the evolution because the evolutionary times become too short and induce numerical instabilities in the code.
        
        Fig.~\ref{fig:PureHe_ts} shows the duration of the He-MS and the resulting mass of the CO core, M$_{\rm CO}$, divided by the M$_{\rm CO}$ of the star with $0\, \sigma$, as a function of the initial mass, M$_{\rm HE-ZAMS}$, and for different reaction rates. Looking at the same initial mass, models computed with higher rates have slightly longer He-MS than models with lower rates. On the other hand, the M$_{\rm CO}$ is almost not affected by the different rates adopted, and the differences are below the 1\% in the worst case (as shown in the central panel). The main difference at the end of the He-MS between models computed with different rates is the carbon-to-oxygen ratio in the CO core, that is shown in the bottom panel. In the case of the 20\Msun\ stars, we find the largest variation, that goes from 1 (i.e., the CO core is composed 50\% of $^{12}$C and 50\% of $^{16}$O) to about 0.01 (i.e., the CO core is almost totally composed of $^{16}$O).
        
        To show the main differences on the evolution of our pure-He stars due to different assumed rates, in Fig.~\ref{fig:PureHe_Kipp} we plot the Kippenhahn diagrams of pure-He stars with M$_{\rm ZAMS} = 40$~\Msun\ with different rates. The mass lost by the star through the evolution is just about 1.2~\Msun, and is manly lost during the He-MS. During the He-MS, these stars build up massive convective cores that mix the star up to the 92\% of the entire mass.
        Different rates slightly affect both the luminosity of the model and the size of the convective core leading to relative differences of He-MS lifetimes up to $~18$\% between the model computed with $+ 3\, \sigma$ and the one with $- 3\, \sigma$ (appreciable also in Fig.~\ref{fig:PureHe_ts}).
        
        After the He-MS, stars are totally radiative and have massive CO cores that do not change much depending on the assumed rate of the \CagO\ reaction, as shown in Fig.~\ref{fig:PureHe_ts}. On the other hand, the carbon-to-oxygen ratio is very different. The carbon (oxygen) mass fraction in the CO core are 0.43, 0.17 and $\approx 0.004$ (0.52, 0.79 and 0.91) for stars with $- 3\, \sigma$, $0 \, \sigma$ and with $+ 3\, \sigma$, respectively. Up to this stage, the evolution of these three models is quite similar.
        
        Stars ignite carbon in their centre when temperatures reach 0.8~--~1~GK. The carbon burning phase lasts for about 21 years in the case of the $- 3\,\sigma$ model, 10 years and 5 years for models with $0 \, \sigma$ and $+ 3\,\sigma$, respectively. During this phase, a helium burning shell forms above the CO core. The neon burning starts when the central temperatures are about 1.4 GK and all the neon in the central parts of the star is depleted in about 46, 6 and 2 days for models computed with $- 3 \, \sigma$, $0\, \sigma$ and $+ 3\, \sigma$, respectively. During this phase, depending on the carbon mass fraction in the CO core, models may turn-on a carbon burning shell above the centre of the star, which develops a big intermediate convective region in the CO core. Such convective region may lead to a growth of the CO core, and provides fresh carbon to the C-burning shell. The energy released by the off-centre carbon burning sustains the external layers of the star, and avoids the stellar collapse due to the pair creation instability. The important role of the convective C-burning shell in preventing the collapse has been reported by other authors \citep[such as][]{Takahashi2018, Farmer2020}.
        
        Finally, oxygen burning takes place at the centre of the star when the central temperature reaches $\approx{}2$ GK. 
        At this point of the evolution, only the model computed with $-3\, \sigma$ is globally stable and we can follow the evolution until oxygen's depletion. The model with the $0 \, \sigma$ becomes dynamically unstable at about half of the oxygen burning, while the one with $+ 3\, \sigma$ becomes  dynamically unstable at the beginning of oxygen burning.
        The large amount of energy released by oxygen burning generates a small convective core of about 1.63~\Msun, 2.87~\Msun\ and 0.83~\Msun\ at the end of the computation for models with $- 3\, \sigma$, $0\, \sigma$ and $+ 3\, \sigma$, respectively. If possible, all the oxygen is depleted leaving a core mainly composed of silicon and sulfur. The model with $- 3\, \sigma$ burns the oxygen in about 30 days. At this point the $-3\, \sigma$ model is near to silicon burning and to the final core collapse.  
        
        Fig.~\ref{fig:PureHe_gamma1} shows the evolution of the \G1{TOT} value of pure-He stars' models with M$_{\rm ZAMS} = 40$~\Msun\ with different rates adopted for the \CagO\ reaction. This figure shows the evolution from the ignition of carbon. As already seen in Fig.~\ref{fig:PureHe_Kipp}, our models start to behave differently after carbon depletion from the core, depending on the amount of carbon-to-oxygen ratio in their CO cores. Models with \fco\ corresponding to \CagO\ rates between $0\, \sigma$ and $+3\, \sigma$ become more and more unstable as they evolve. In contrast, models with $-1\, \sigma$, $-2\, \sigma$ and $-3\, \sigma$ tend to stabilize and continue their evolution until oxygen depletion, avoiding PI.

    \subsection{Pre-supernova masses}
    \label{sec:Finmass_He}
        Table~\ref{tab:HeTable} in Appendix~\ref{appendix:Tab} shows the properties of the pure-He stars computed in this work. The final fate is assigned as follows: when $\G1{TOT-}>0$, the star is stable, continues to evolve through the last burning phases and a final core collapse (CC) will form a black hole with mass $\leq{}M_{\rm Pre}$. When $\G1{TOT-}\leq{}0$, the star is dynamically unstable. Depending on its mass, it may start to pulsate losing mass through PPI, or it might explode after a single pulse, becoming a PISN and leaving no remnant. In Table~\ref{tab:HeTable}, we refer to both PPI and PISN as PI. The evolution through such phases must be followed by means of hydro-dynamical simulations, as done by, e.g.,  \citet{Woosley2017}, \citet{Takahashi2018} and \citet{Marchant2019}. Since we did not run hydro-dynamical simulations, we conservatively assume that both stars going through PPI and stars undergoing PISN leave no compact objects. Hence, we will obtain lower limits to the maximum compact object mass.
        
        In Appendix \ref{appendix:Mesa}, we show a comparison between our pure-He star models and those computed by \citet{Farmer2020} with the \textsc{mesa} stellar evolutionary code \citep{paxton2011,paxton2013,paxton2015,paxton2018}. From such a comparison we see that our stable models at the end of the computation are very similar to the models of \citet{Farmer2020} that are dynamically stable during the oxygen burning phase, which evolve directly toward the final CC.
        
        Due to both the low metallicity adopted, that leads to mass-loss rates of $\approx 5 \times 10^{-6}$~\Msun yr$^{-1}$, and to the short lifetimes of pure-He stars, of the order of $10^5$ yr, all the pure-He models computed have pre-supernova masses, M$_{\rm Pre}$, very close to $M_{\rm He-ZAMS}$.
        As discussed before and also shown in Fig.~\ref{fig:PureHe_ts}, different rates for the \CagO\ reaction change the carbon-to-oxygen mass fraction at the end of the He-MS, but do not influence the He and CO core mass.
        
\begin{figure}
            \includegraphics[width=0.48\textwidth]{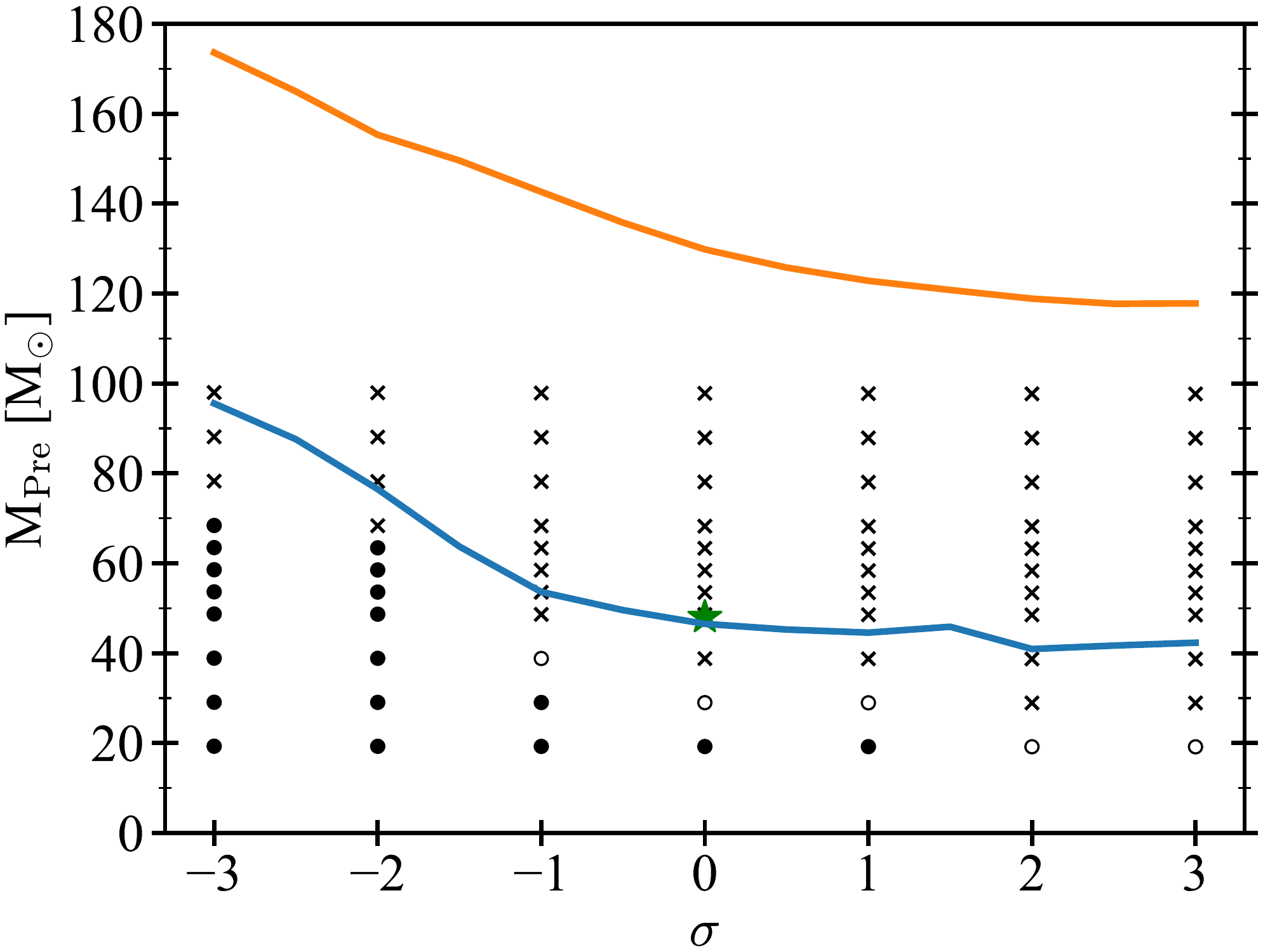}
            \caption{Pre-Supernova masses of pure-He stars as a function of the \CagO\ reaction rate. Circles indicate models that are dynamically stable while burning oxygen. Empty circles: models with a central oxygen mass fraction between 0.3 and 0.2; filled circles: models with a central oxygen mass fraction lower than 0.2. Black crosses indicate models that are dynamically unstable (i.e. \G1{TOT-}\ < 0). Table~\ref{tab:HeTable} lists all the values plotted here. Continuous blue and orange lines indicate the lower and higher mass gap edges by \citet{Farmer2020}, respectively. The green star at $0\, \sigma$ indicate the $M_{\rm Gap}$ = 48~\Msun\ found by \citet{Woosley2017} for pure-He stars.
                    }
            \label{fig:PureHe_Farmer}
        \end{figure}

        Fig.~\ref{fig:PureHe_Farmer} shows the lower edge of the PI gap as a function of the \CagO\ reaction rate. For pure-He stars, the maximum He-ZAMS mass that results in a stable model and evolves until core collapse is about 68~\Msun\ and 19~\Msun, in the case of \CagO\ reaction rates $-3\, \sigma$ and $+3\, \sigma$, respectively.
        This trend is similar to the one shown by \citet{Farmer2020}, but in our case the maximum mass is always lower by $\approx{}10-20$~\Msun\ than theirs, for each assumed rate. This happens because, on the one hand, {\sc parsec} includes slightly different physical prescriptions for the computation of the evolutionary tracks (e.g., for the opacity, mass-loss and convection treatment) and, on the other hand, we do not compute the hydro-dynamical evolution of the models after they become unstable. Hence,  we conservatively assume that all stars with \G1{-} $\leq$ 0 leave no compact object, because we cannot distinguish between PPI and PISN and we cannot model mass loss during PPI. Hydro-dynamical simulations of PPI \citep{Marchant2019,Farmer2020} show that, depending on the mass of the star, the pulses may eject just few tenths of \Msun\ before the star stabilizes. After this small mass-loss, the star continues its evolution until core collapse. Hence, including hydrodynamical evolution might stabilize some of our models and lead to a higher value of the $M_{\rm gap}$. For the same reason, our maximum BH mass obtained adopting the standard \CagO\ rate differs by $\approx{}20$~\Msun\ with respect to the maximum BH mass obtained from pure-He stars by \citet{Woosley2017}, that is 48~\Msun, and by \citet{Leung2019}, 50~\Msun.

        The analysis of pure-helium stars, the comparison with \citet{Farmer2020} pure-He models (Appendix~\ref{appendix:Mesa}), and the fact that our results are comparable to previous work confirm that our criterion for stability is a conservative one to decide whether a stellar model is stable, even if we do not follow the hydrodynamical evolution of the final stages.
        Since we do not follow the evolution through the pulsation PPI phase, our results are lower limits to the $M_{\rm gap}$.

%
%
\section{Full stellar models with hydrogen envelope}
\label{sec:Results_H}
    
    We now consider stellar models with hydrogen envelopes, computed with the same version of {\sc parsec} as described in Section~\ref{sec:Tracks_He}.
    Grids of evolutionary tracks are calculated with different values of the multiplier parameter, \fco, as shown in Table~\ref{tab:reacrate}. 

    \subsection{Evolutionary tracks}
    \label{sec:Tracks}
    %
            %
        \begin{figure}
            \includegraphics[width=0.48\textwidth]{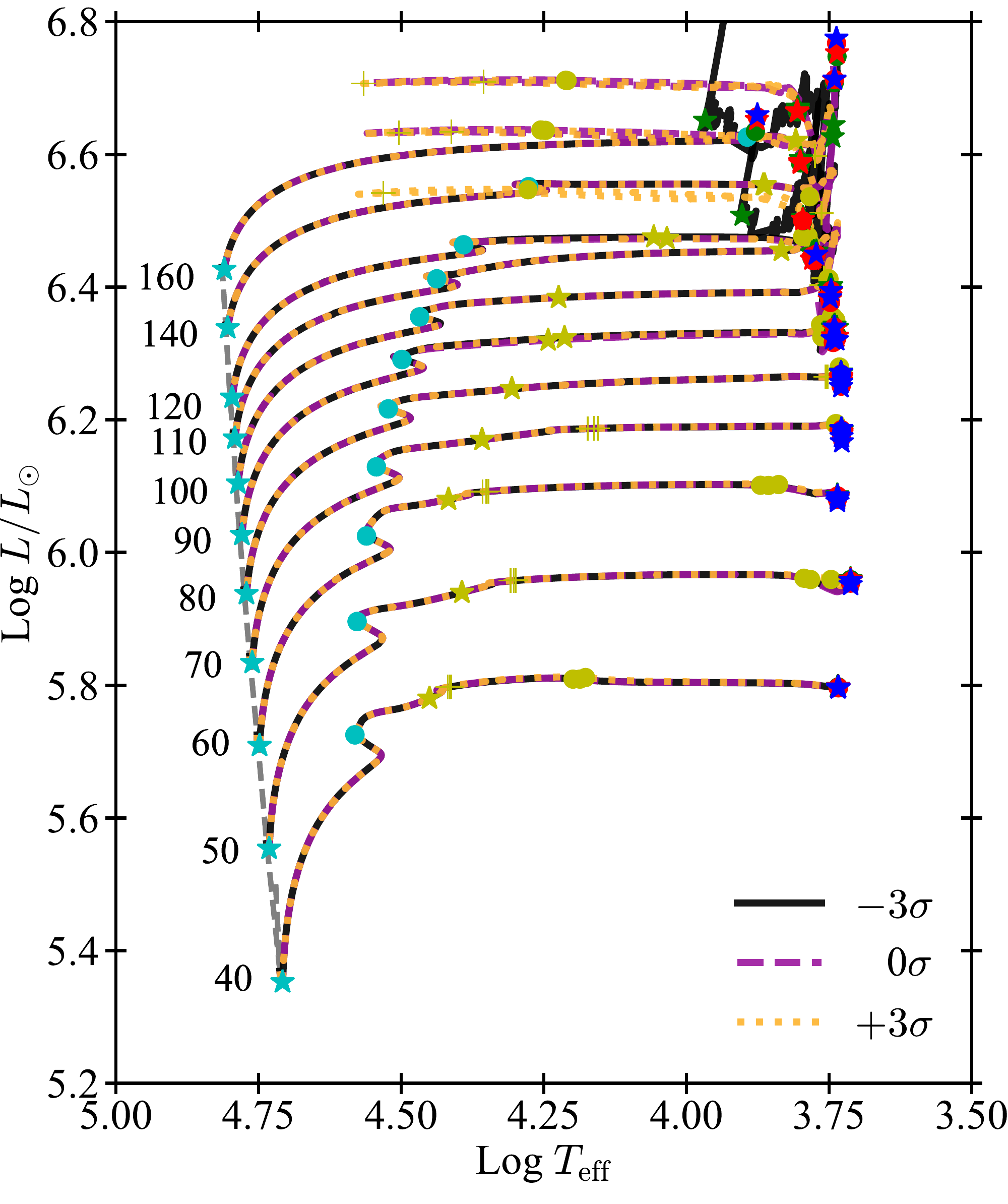}
            \caption{HR diagram of the three sets of H stars computed with $Z = 0.0003$ and with different reaction rates. The colour code of lines and symbols is the same as in Fig.~\ref{fig:PureHe_HR}. The cyan stars (circles) indicate the beginning (the end) of the hydrogen burning phase. The grey dashed line is the ZAMS. The numbers on the left of the ZAMS are the values of M$_{\rm ZAMS}$  in units of \Msun.
                    }
            \label{fig:HR}
        \end{figure}
        %
        \begin{figure}
            \includegraphics[width=0.5\textwidth]{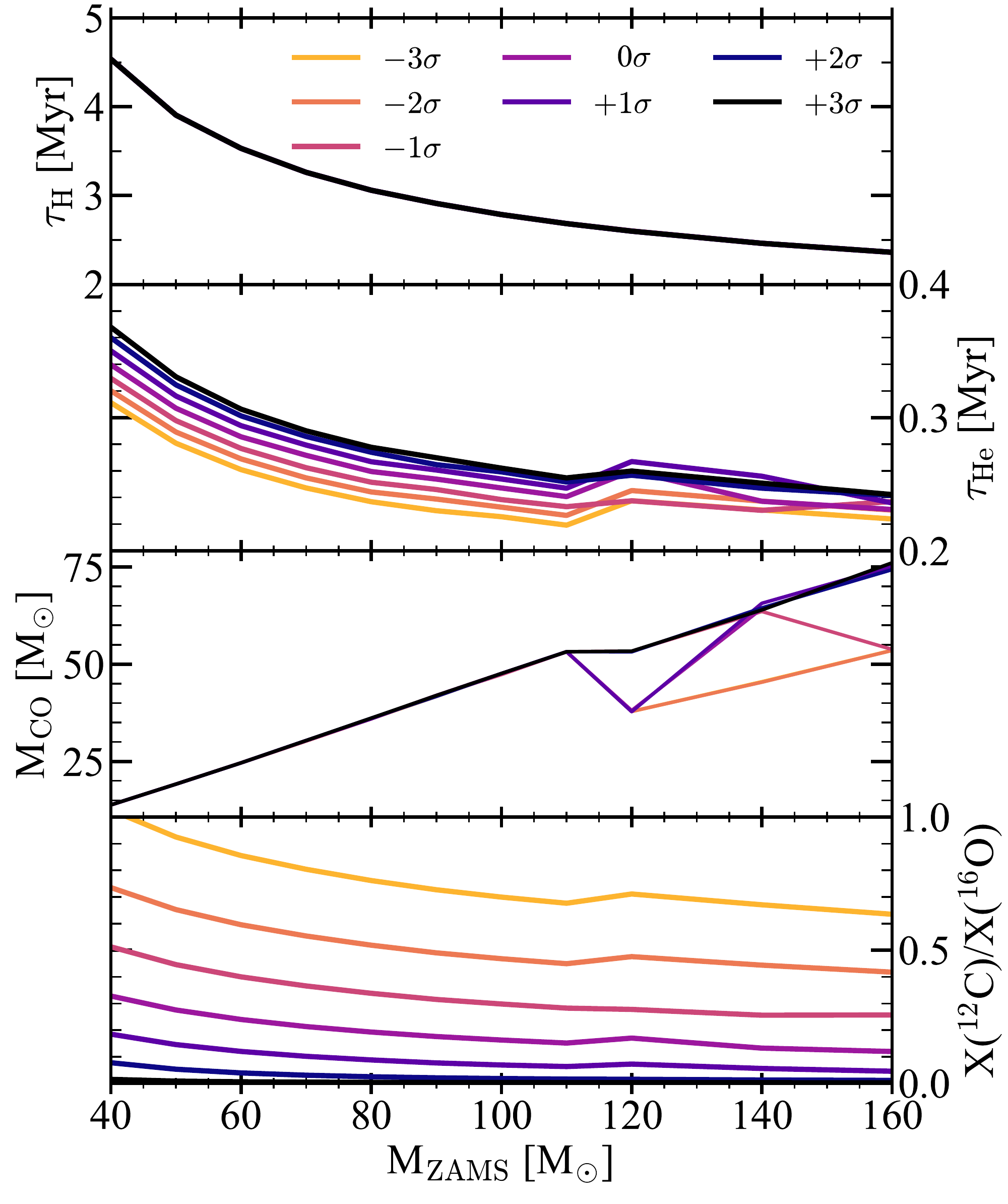}
            \caption{Same as Fig.~\ref{fig:PureHe_ts}, but for stars with hydrogen envelopes. The upper panel shows the MS lifetimes of hydrogen stars versus the ZAMS mass. The second panel from the top shows the CHeB lifetimes. The third panel shows the CO core mass. 
            Different colours indicate different \CagO{} rates (several of these lines overlap).  Table~\ref{tab:HTable} lists all the relevant quantities.
            The lower panel shows the carbon-to-oxygen ratio in the CO cores at the end of CHeB.
            }
            \label{fig:H_ts}
        \end{figure}

        Each grid contains stellar models with initial masses of 40, 50, 60, 70, 75, 80, 85, 90, 95, 100, 105, 110, 120, 140 and 160~\Msun. Our models begin their evolution from the ZAMS and evolve until the end of the central oxygen burning, when possible. 
        The initial metallicity and helium content are $Z=0.0003$ and $Y=0.249$, respectively \citep{Bressan2012}. 

        In Fig.~\ref{fig:HR},  we plot the HR diagram of three selected sets of tracks computed with \fco\ corresponding to $+3\, \sigma$, $0\, \sigma$, $-3\, \sigma$. 
        The \CagO\ reaction does not affect the hydrogen burning phase and the models with the same mass evolve in the same way during the MS. 
        
        After the MS, the stellar cores rapidly contract until their central temperature is high enough for helium ignition. Stars cross the so called Hertzsprung gap while the effective temperatures decrease. Whether a massive star ignites helium as blue super giant (BSG) or as a yellow/red super giant (YSG/RSG) star depends on many stellar and physical parameters (such as the stellar mass, the metallicity, the mass-loss, the opacity and previous mixing efficiency).
        In our sample, all models with M$_{\rm ZAMS}~\leq~90$~\Msun\ ignite helium as BSG stars, while more massive ones ignite helium as YSG/RSG stars. 
        
        During the CHeB, stars develop a convective core that stops growing when the central He mass fraction falls below $\approx{}0.5$. In this phase, the carbon fraction initially increases up to a maximum and then decreases because of the effects of the \CagO\ reaction. As the \CagO\ rate increases (from $- 3\, \sigma$ to $+ 3\, \sigma$), more carbon is converted into oxygen, eventually leaving an almost carbon-free core at the end of the CHeB phase in the case of the model computed with $+ 3 \, \sigma$. However, these differences do not affect much the CHeB lifetimes and the mass of the CO core, as shown in Fig.~\ref{fig:H_ts}. During the CHeB, the star develops several intermediate convective regions in the envelope, that remain detached from the He core for the remaining evolution. 
        As shown in Fig.~\ref{fig:H_ts}, the CHeB lifetimes are only slightly affected by the assumption of different \fco\ values, with a maximum difference of $\approx 5 \times 10^4$ yrs in the case of the M$_{\rm ZAMS} = 40$~\Msun\ star. Moreover, adopting different \fco\ values has no impact on the CO  core masses left after the CHeB phase.
        Only in stars with M$_{\rm ZAMS} \geq 120$~\Msun\ the \CagO\ rate does affect the CO mass, because of the presence of an efficient envelope undershooting, as it will be further discussed in Section~\ref{sec:env_ov}.
        Fig. \ref{fig:H_ts} shows that the major effect of changing \fco\ is on the central carbon-to-oxygen ratio left at the end of the CHeB, as in pure-He stars.
 
        At such low metallicity, all our models evolve through advanced burning phases as RSG stars. This is due to the low stellar winds that allow stars to keep their massive H-rich envelope.
        
        After the CHeB phase, the core starts to contract again increasing its density and temperature. Meanwhile, neutrino losses tend to cool down the core, but, when the central temperature reaches $\approx 1$ GK, carbon is ignited. 
        Stars computed with lower \CagO\ rates have longer lifetimes than models computed with higher \CagO\ rates.
        
        After carbon depletion, neon photo-disintegration begins at the stellar centre when the temperature rises up to about 1.2 -- 1.5 GK. During carbon and neon burning, our models have radiative cores and convective envelopes. As already seen for pure-He stars, during the central neon burning phase, carbon burns off-centre in a shell that, in case of low \CagO\ rates, is able to turn on a intermediate convective region. 
        
        This convective zone inside the CO core continuously provides fresh carbon to the burning shell that sustains the overlying layers, thus delaying the ignition of oxygen in the core.
        As the evolution proceeds further, \G1{TOT} may become lower than 4/3 + 0.01, indicating that the star becomes globally unstable. In this case, the computation is stopped. Otherwise, we continue the evolution until the end of central oxygen burning, when eventually the code encounters some convergence problems. Apart from a few isolated cases, the values of the final central oxygen mass fractions in stable models are between 0.01 and 0.3  (Table \ref{tab:HTable}).

        After the central oxygen burning phase, the evolution of the core and of the envelope become decoupled. Therefore, in analogy with the pure-He models, we can assume that if the star remains dynamically stable during the core oxygen burning, it will evolve towards the advanced silicon and iron burning phases and finally, to the core collapse. 

\begin{figure*}
            \includegraphics[width=0.33\textwidth]{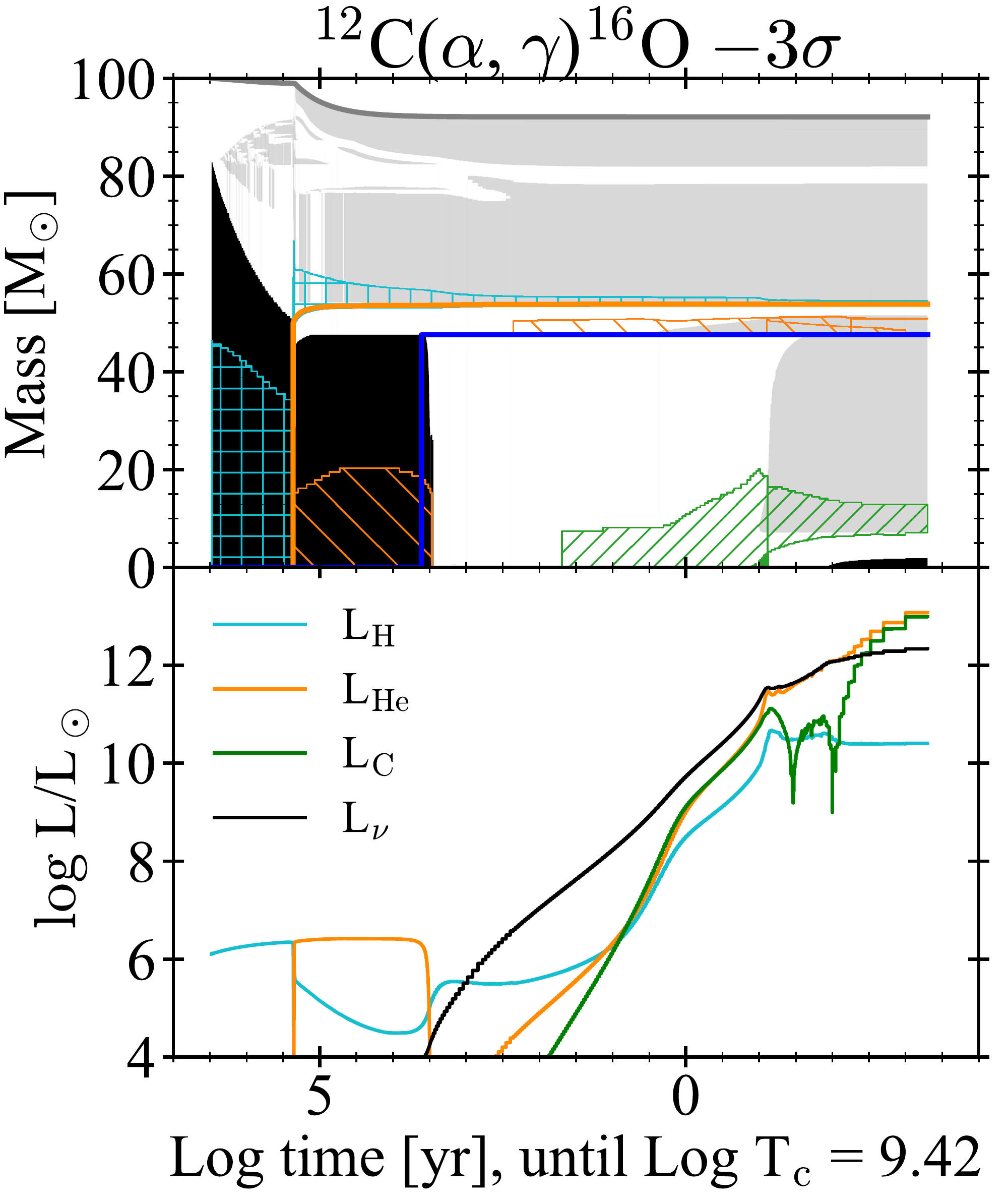}
            \includegraphics[width=0.33\textwidth]{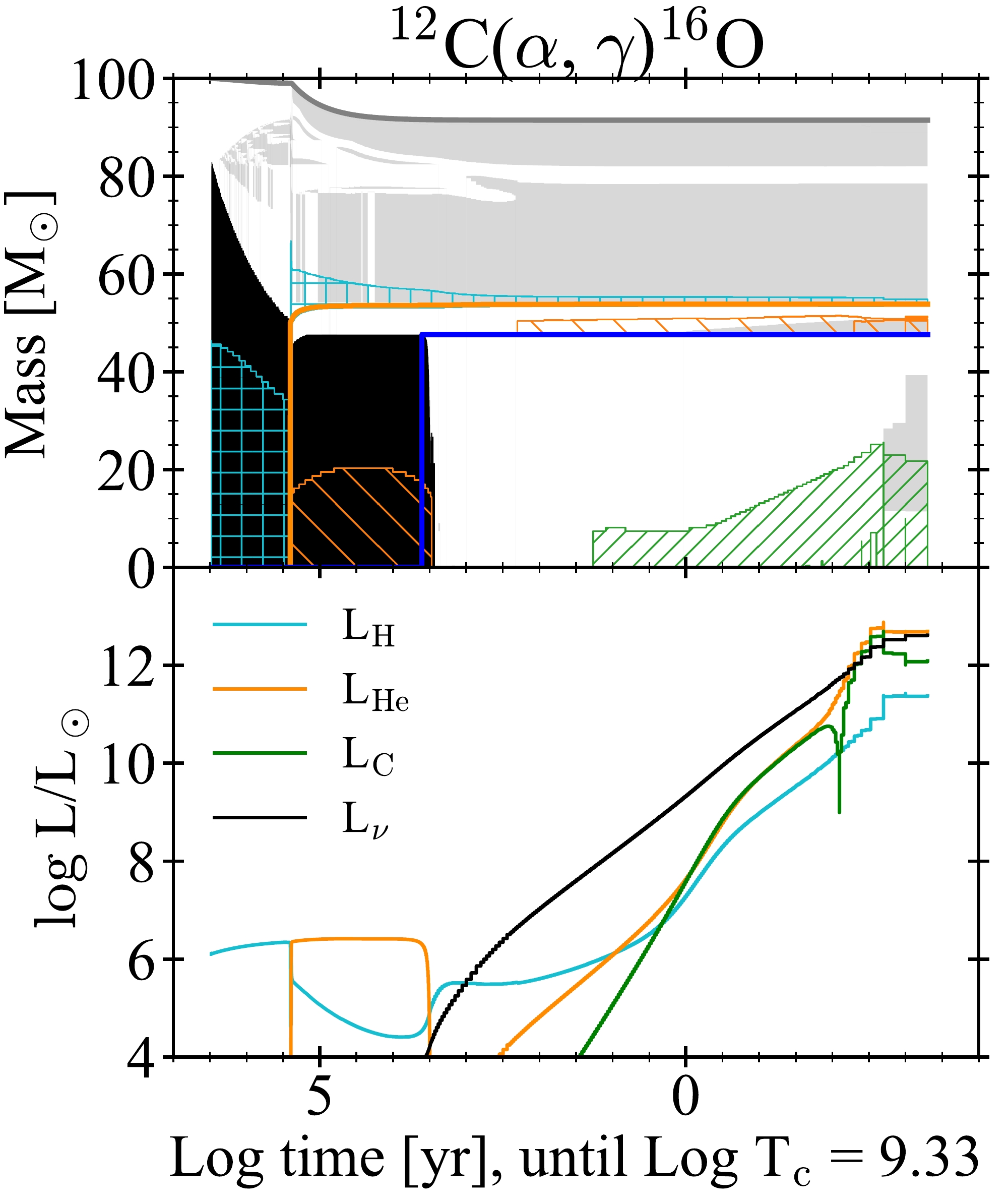}
            \includegraphics[width=0.33\textwidth]{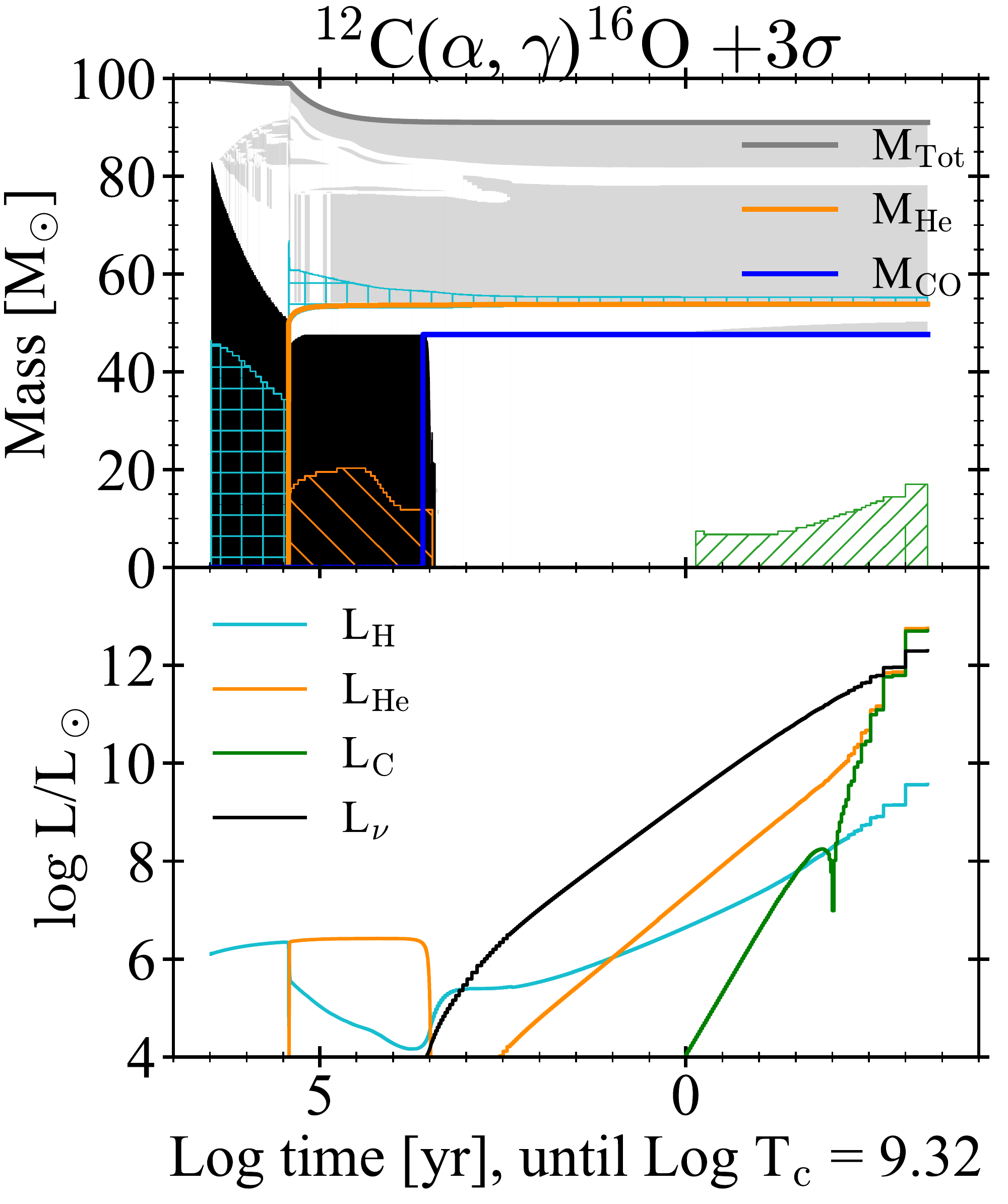}

            \caption{Upper panels: Same as Fig.~\ref{fig:PureHe_Kipp}, but for hydrogen stars with M$_{\rm ZAMS}=100$~\Msun. The continuous gray lines indicate the total stellar mass. The cyan coloured areas indicate the hydrogen burning regions, which contribute at least for the 1\% of the
            L$_\mathrm{H}$.
            Lower panels: Luminosity as a function of time from H-burning, He-burning, C-burning and neutrinos in cyan, orange, green and black, respectively.
            }
            \label{fig:H100_Kipp}
        \end{figure*}

        Fig.~\ref{fig:H100_Kipp} shows the Kippenhahn diagrams of stars with M$_{\rm ZAMS} = $~100~\Msun\ computed with \fco\ corresponding to \CagO\ rate $+ 3\, \sigma$, $0 \, \sigma$ and $- 3\, \sigma$. The three models evolve in a similar way during the MS and the CHeB phases. They build up a convective core that decreases over time in the MS. At the end of the MS, stars have a He core of $\approx 50$~\Msun. During the CHeB, the models develop a convective core of $\approx{}47$~\Msun, in which all the helium is converted to carbon, oxygen and neon.
        
        When temperature and density are high enough, carbon is ignited in the core. It burns for 12 yr, 2.8 yr and only 4 months in the case of models computed with \fco\ corresponding to $- 3\, \sigma$, 0 and $+ 3 \, \sigma$, respectively. After the carbon burning, neon photo-disintegration takes place in the star centre, and it lasts for about 20, 2 and 1.4 days for models computed with $- 3\, \sigma$, 0 and  $+ 3\, \sigma$, respectively. 
        At this point, only the model computed with $- 3\, \sigma$ remains globally stable and burns central oxygen in a small convective core of about 2~\Msun. At the end of the computation, this 2~\Msun\ core is composed of 15\% $^{16}$O, 45\% $^{28}$Si, 30\% $^{32}$S and 10\%  heavier elements. 
        This star will continue its evolution through the most advanced burning phases and then will collapse forming a BH. The other models computed with $0\, \sigma$ and $+ 3\, \sigma$ do not have sufficient $^{12}$C in the CO core to prevent PI, hence they become unstable before the ignition of oxygen.
        
\begin{figure}
            \includegraphics[width=0.48\textwidth]{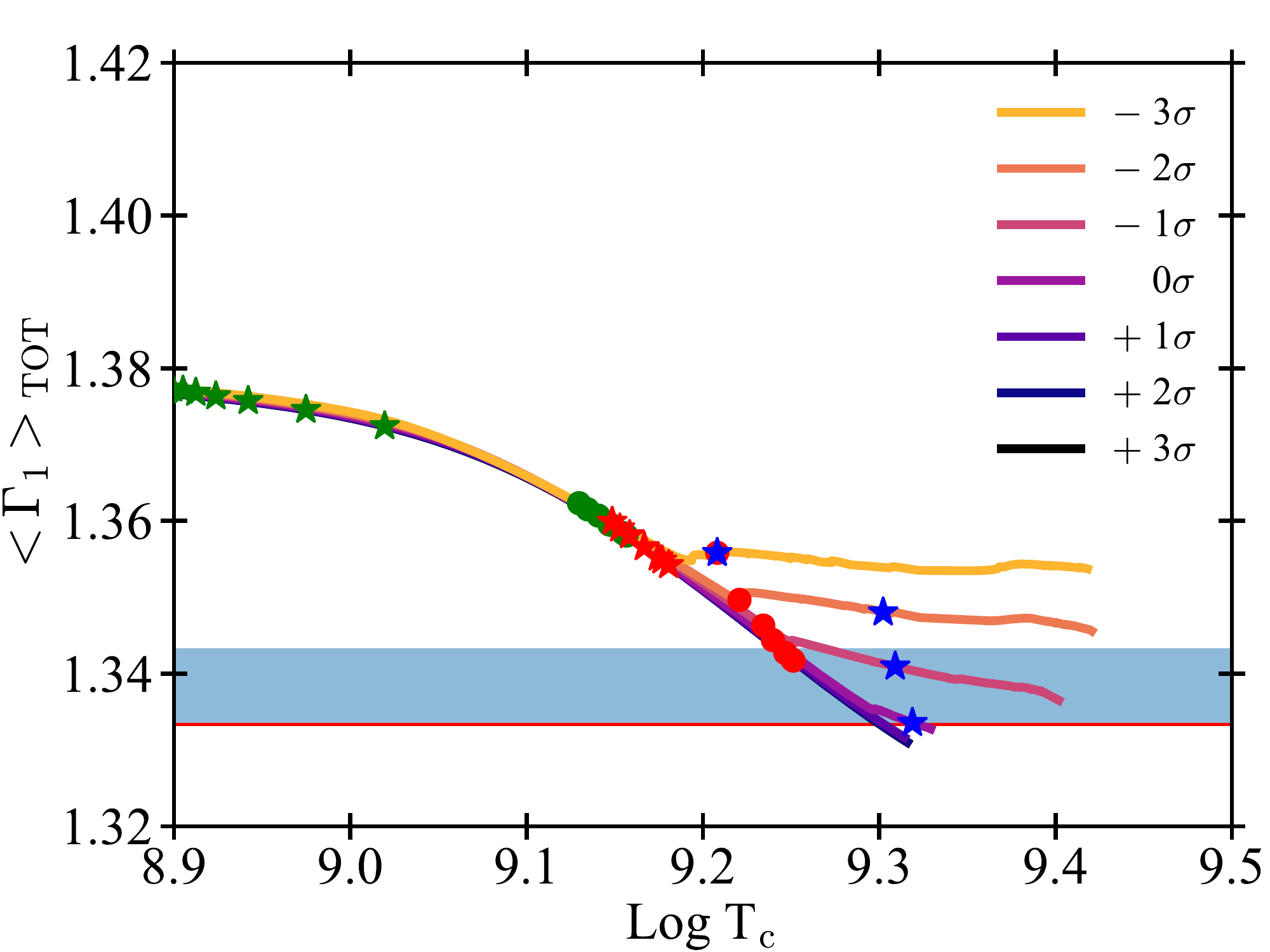}
            \caption{Comparison of the \G1{TOT} values versus the central temperature of stars with hydrogen envelopes and M$_{\rm ZAMS}$ = 100~\Msun. Here, we use the same colour code for lines and symbols as in Fig.~\ref{fig:PureHe_gamma1}. The lines corresponding to +2 and $+3\, \sigma$ are barely visible, because they almost perfectly overlap with the $+1\, \sigma$ line.
            }
            \label{fig:H100_gamma1}
        \end{figure}

        Fig.~\ref{fig:H100_gamma1} shows the evolution of the averaged first adiabatic exponent, \G1{TOT}, of the models with 100~\Msun\ and computed with different \CagO\ rates. 
        Depending on the assumed rate the star enters or not in the unstable regime. In this case, only models with $-3\, \sigma$ and $-2 \, \sigma$ remain stable during the oxygen burning phase. The other models become unstable shortly after the end of the neon burning phase.

        %

    \subsection{Pre-supernova masses}
    \label{sec:Finmass_H}

        Table~\ref{tab:HTable} in Appendix~\ref{appendix:Tab} shows the results for the stars with H envelopes. The columns list the same quantities as shown in Table~\ref{tab:HeTable}, with the exception of the first adiabatic exponent (defined in Section~\ref{sec:dyn_inst}), that here is computed in two ways: i) for whole star, \G1{TOT-} = \G1{TOT} - (4/3 + 0.01), and ii) for the He core, \G1{Core-} = \G1{Core} - (4/3 + 0.01). We find that \G1{Core} is always slightly lower than \G1{TOT}. The envelope of massive stars could be near the dynamical instability too, depending on the ionization state of the plasma \citep{Stothers1999}. \G1{Core} and \G1{TOT} are not significantly different, because \G1{} is a weighted integral and the external parts of the envelope contribute less than the internal parts of the star.
        We use the same methodology adopted for the pure-He stars to assign the final fate of stars. In the case of \G1{TOT-} > 0, the star is considered globally stable and  evolves until the final CC. On the other hand, if \G1{TOT-} $\leq$ 0, the star becomes dynamically unstable and the PI-induced collapse leads to a PPI or a PISN. This fate is indicated with the label PI in the Table. We find that the final fate does not change if we assume \G1{TOT} or \G1{Core}.
        
\begin{figure}
            \includegraphics[width=0.48\textwidth]{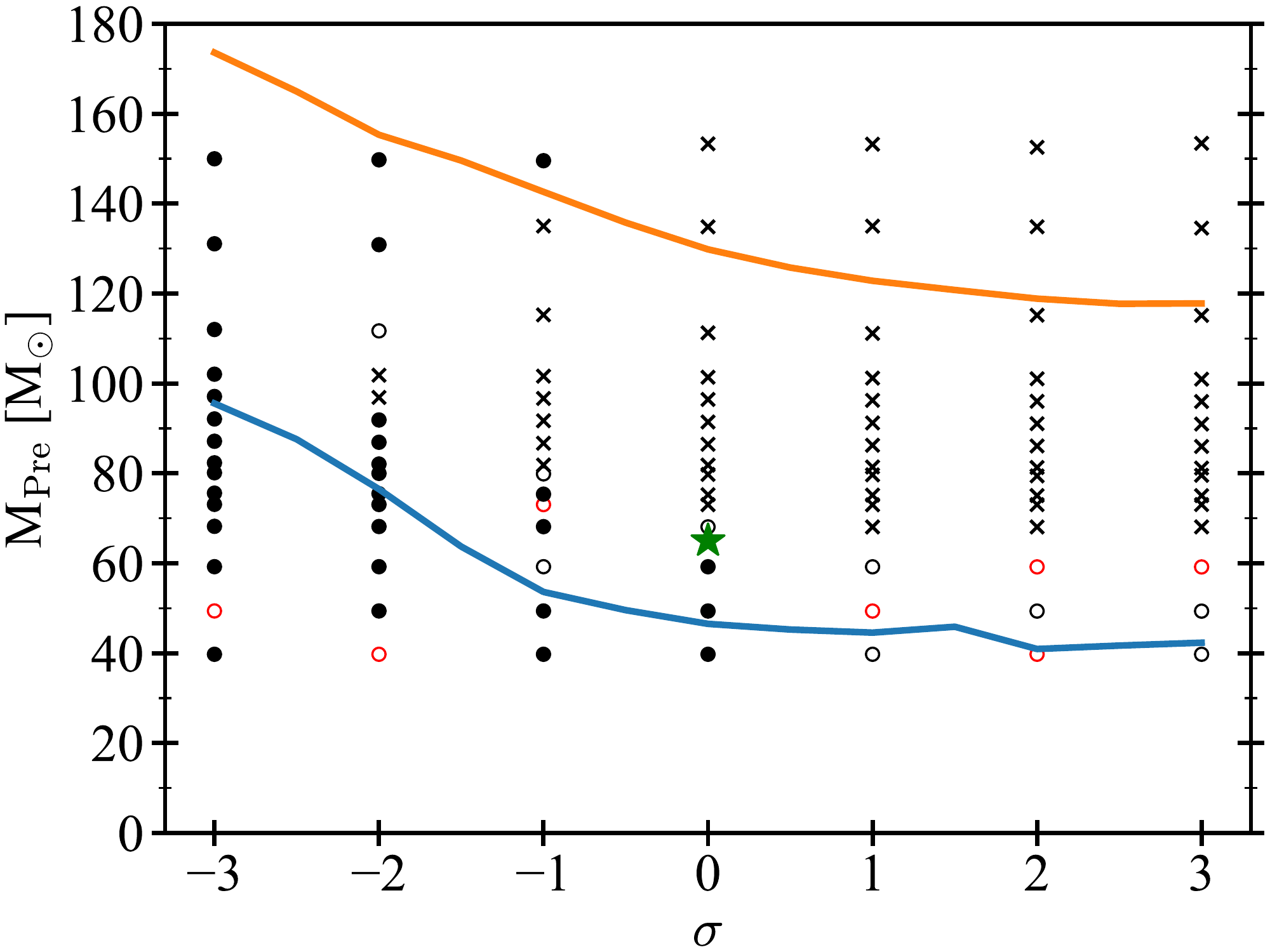}
            \caption{Pre-supernova masses as a function of the \CagO\ reaction rate for stars with H envelopes. 
            Circles indicate models that are dynamically stable while burning oxygen. Red empty circles: models with a central oxygen mass fraction higher than 0.3. Black empty circles: models with a central oxygen mass fraction between 0.3 and 0.2; filled circles: models with a central oxygen mass fraction lower than 0.2. Black crosses indicate models that are dynamically unstable (i.e. \G1{TOT-}\ < 0). 
            The PI mass gap is completely filled in the  $-3\, \sigma$ case, because of the effect of the dredge-up (see Section~\ref{sec:Finmass_H} for details). Continuous blue and orange lines indicate the lower and higher mass gap edges by \citet{Farmer2020}, respectively. The green star at $0\, \sigma$ indicates the $M_{\rm gap}$ = 65~\Msun\ found by \citet{Woosley2017} for stars with hydrogen envelope.
                    }
            \label{fig:H_Farmer}
        \end{figure}

        Fig.~\ref{fig:H_Farmer} shows the pre-supernova masses of stars with H envelope as a function of the \CagO\ rate. 
        The lower edge of the mass gap ($M_{\rm gap}$) ranges from $\approx$ 60~\Msun\ to $\approx$ 92~\Msun\ in the case of models computed with $+ 3\, \sigma$ and $-2\, \sigma$, respectively. Interestingly, we find no mass gap in the case of models computed with $- 3\, \sigma$. The maximum M$_{\rm Pre}$ is $\approx$150~\Msun.
        Moreover, we find that stars computed with $-2\, \sigma$ and with initial masses of 140~\Msun\ and 160~\Msun\ do not become unstable and burn oxygen non-explosively, with final M$_{\rm Pre}$ masses of~$\approx 131$~\Msun\ and $\approx 150$~\Msun\,respectively.
        
        The disappearance of the mass gap for models with $-3\, \sigma$ is due to a dredge-up during the CHeB, that reduces the mass of the CO core with the effect of stabilizing the core. This process is described in detail below. 
        
        In this case, the lower edge of the PI window is higher than that found by \citet{Farmer2020}, because we consider stars that still have their hydrogen envelopes.
        The range of maximum masses we find is in agreement with the maximum BH mass $\approx{}65$~\Msun, found by \cite{Woosley2017} for stars with hydrogen envelope, and overlaps with the range obtained by \citet[][]{Takahashi2018}, that is between $100$ and $270$~\Msun.
        
    \subsection{
    Envelope undershooting and dredge-up during the advanced phases}
        \label{sec:env_ov}
        
        \begin{figure*}
            \includegraphics[width=0.48\textwidth]{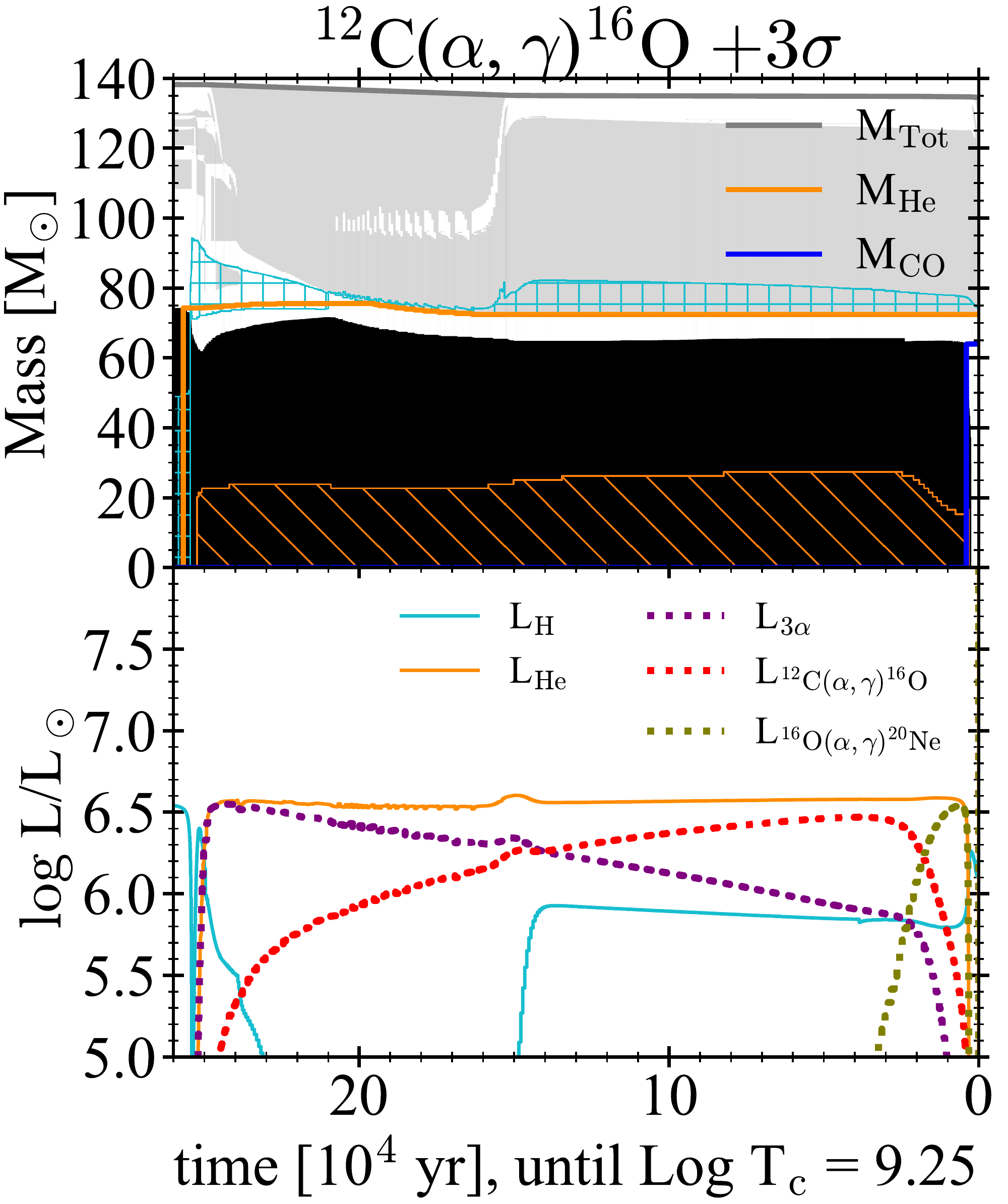}
            \includegraphics[width=0.48\textwidth]{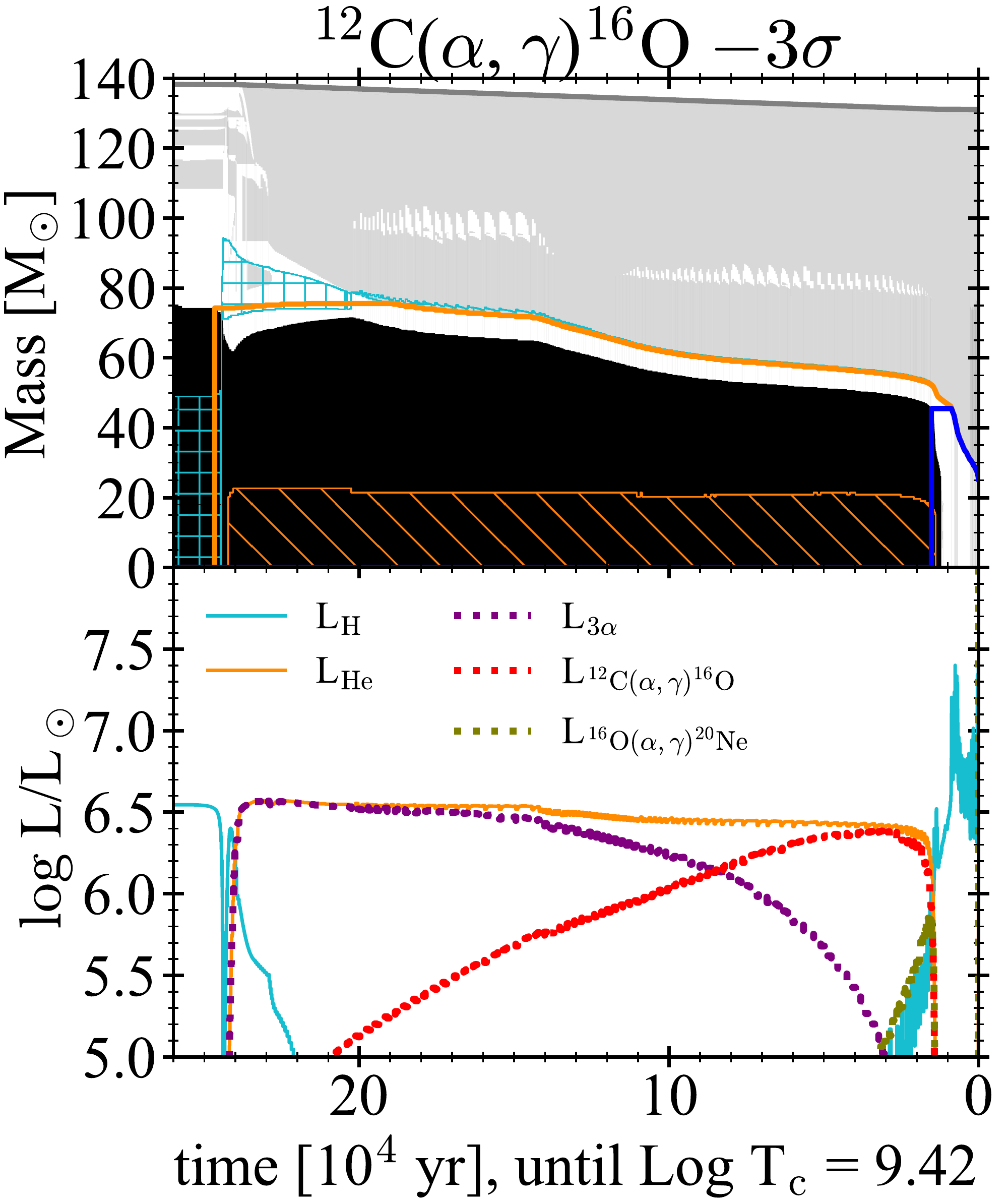}

            \caption{
            Left-hand panels: same as Fig.\ref{fig:H100_Kipp}, but the evolution is zoomed on the CHeB phase of the M$_{\rm ZAMS}=140$ \Msun\ star computed with the \CagO\ reaction rate $+3\, \sigma$. Time until the final collapse is in linear scale. In the lower panel the dotted lines show the contribution to the He luminosity from the CHeB phase main reactions, the triple-$\alpha$, the \CagO\ and the $^{16}$O($\alpha$, $\gamma$)$^{20}$Ne in purple, red and green, respectively.
            Right-hand panels: Same plots as those shown in the left-hand panels, but for a model with M$_{\rm ZAMS}=140$ \Msun\ and $- 3\, \sigma$.
            }
            \label{fig:H140_KippZOOM}
        \end{figure*}
        
        \begin{figure}
            \includegraphics[width=0.48\textwidth]{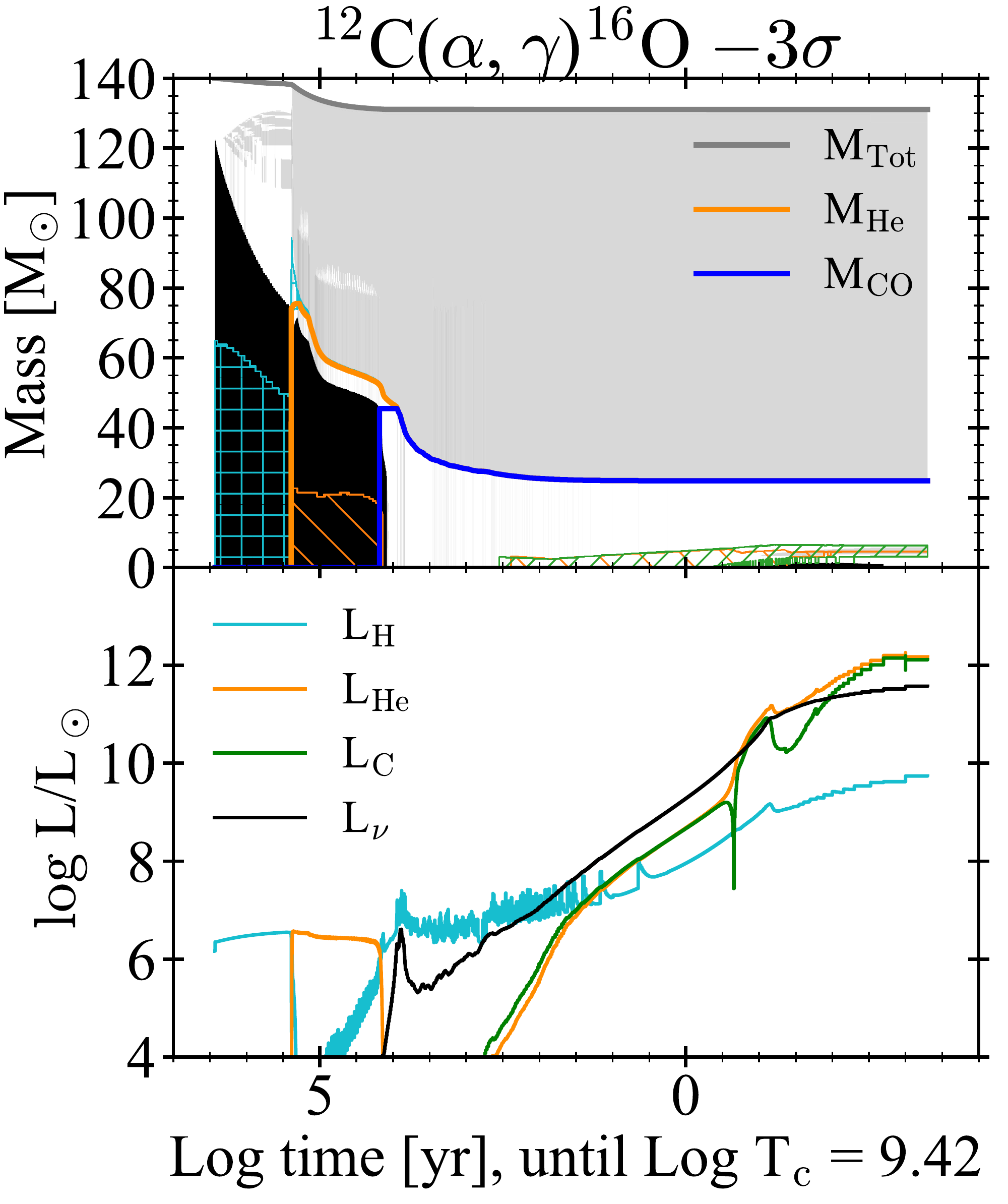}

            \caption{Same as Fig.\ref{fig:H100_Kipp}, but for the M$_{\rm ZAMS}=140$ \Msun\ star with $- 3\, \sigma$. 
            }
            \label{fig:H140m3sig_Kipp}
        \end{figure}
        
        The dredge-up is an opacity-driven process that begins when the star approaches the location of RSG. The envelope expands and cools, the opacity rises and the radiative temperature gradient becomes superadiabatic in the inner envelope layers.
        As a consequence, the base of the convective envelope extends inwards, crossing the H--He discontinuity, and deepening into the He core (envelope undershooting). 
       
        As described in Sec.~\ref{sec:Tracks_He}, in {\sc parsec} we adopt the undershooting at the bottom of the outermost convective region. As a consequence, only models that develop fully convective envelopes, extending from the surface to the He core, undergo the dredge-up.

        Models with M$_\mathrm{ZAMS} < 120$~\Msun, ignite and deplete helium as BSG/YSG stars (Fig.~\ref{fig:HR}), regardless of the \CagO\ reaction rate assumed. Such models show a complex structure with many intermediate convective regions: they do not develop fully convective envelopes and avoid the dredge-up (see Fig. \ref{fig:H100_Kipp}).

        In contrast, all our models with M$_\mathrm{ZAMS}\geq$ 120~\Msun\  become RSG stars after the MS, regardless of their \CagO{} rate. They develop a large convective envelope and undergo a dredge-up after the ignition of helium, which extracts a few \Msun\ of material from the core (Fig.~\ref{fig:H140_KippZOOM}).
        
        In the following stages, models with high \CagO\ rates have a very different evolution from models with low \CagO\ rates (Fig.~\ref{fig:H140_KippZOOM}). 
        Models with high \CagO\ rates tend to perform blue loops, as in the case of less massive stars \citep{Iben1974, Bertelli1985, Brunish90}. 
        For example, the 140 \Msun\ model with $+3\, \sigma$ \CagO{} rate begins the blue loop when the \CagO\ luminosity exceeds the triple-$\alpha$ one. At higher effective temperatures, the H-burning shell re-ignites and the convective envelope becomes thinner, while some detached intermediate convective zones appear (left-hand panels of Fig.~\ref{fig:H140_KippZOOM}). 
        Envelope undershooting is not applied to such intermediate convective regions. Thus, the penetration into the He-core is not boosted even if, sometimes, convection extends down to the bottom of the H-rich envelope. We will investigate these cases in a follow-up study.
      
        In contrast, models computed with low \CagO\ reaction rates deplete He in the RSG branch, maintaining their large convective envelopes. Hence, the dredge-up process continues (right-hand panels of Fig.~\ref{fig:H140_KippZOOM}), leading to a significant decrease of the He core mass during CHeB. This keeps these models out of the PI regime.
        
        Fig.~\ref{fig:H140m3sig_Kipp} shows the complete Kippenhahn diagram and the luminosity evolution until the end of the O-burning of a star with M$_{\rm ZAMS}=140$~\Msun\ and \CagO\ rate $-3\, \sigma$. 
        During the MS, the star has a big convective core, which decreases over time, and a radiative envelope. As the core recedes, small intermediate convective zones form in the envelope. At the end of the MS, the star moves towards the red part of the HR-diagram expanding and cooling. 
        At the beginning of the CHeB, the star develops a large convective region that, after about 5$\times 10^4$ yrs, extends from the stellar surface to the bottom of the H-rich envelope (upper-right panel of Fig.~\ref{fig:H140_KippZOOM}).
        During this phase, the H-burning shell quenches out and the undershooting at the base of the envelope penetrates in the He core bringing helium up to the surface (dredge-up).

        During CHeB phase, more than $\approx{}20$~\Msun\ of helium are extracted from the core (Table~\ref{tab:HTable}). The temperature at the bottom of the H-burning shell (T$_\mathrm{b}$) reaches 40 Mk.
        At the end of this stage, the envelope is He-rich (with an helium mass fraction of 0.52) and has super-solar metallicity (Z$_{CNO}\approx{0.13}$).
        
        After CHeB, the core contracts and the H-shell ignites again, being the only nuclear energy source for a while. The excess luminosity causes a further quick penetration of the external convection corresponding to a second dredge-up. This brings the H discontinuity very close to the CO core leaving only a very thin He shell.  
        In the meantime, the core contracts and both neutrino and hydrogen luminosities rise. The latter causes a further expansion of the external envelope accompanied by a continuous convective envelope undershooting into the CO core. T$_\mathrm{b}$ reaches about 70 MK. The undershooting mixing and subsequent hot bottom burning continue until carbon ignites in the centre, increasing appreciably the surface metallicity. At this stage, the convective envelope has a mass of $\approx{}105$~\Msun\ and a composition of 3.5\% $^1$H, 23\%  $^4$He, 48\% $^{14}$N, 23\%  $^{16}$O and 2.5\% of $^{20}$Ne and heavier elements. At the end of the central carbon burning phase, lasting for 5.7$\times 10^2$ yrs, the CO core mass is $\approx$24.8~\Msun.
        Subsequent central neon burning lasts for about 2 months, while carbon burns in a shell above the core.
        
\begin{figure}
            \includegraphics[width=0.48\textwidth]{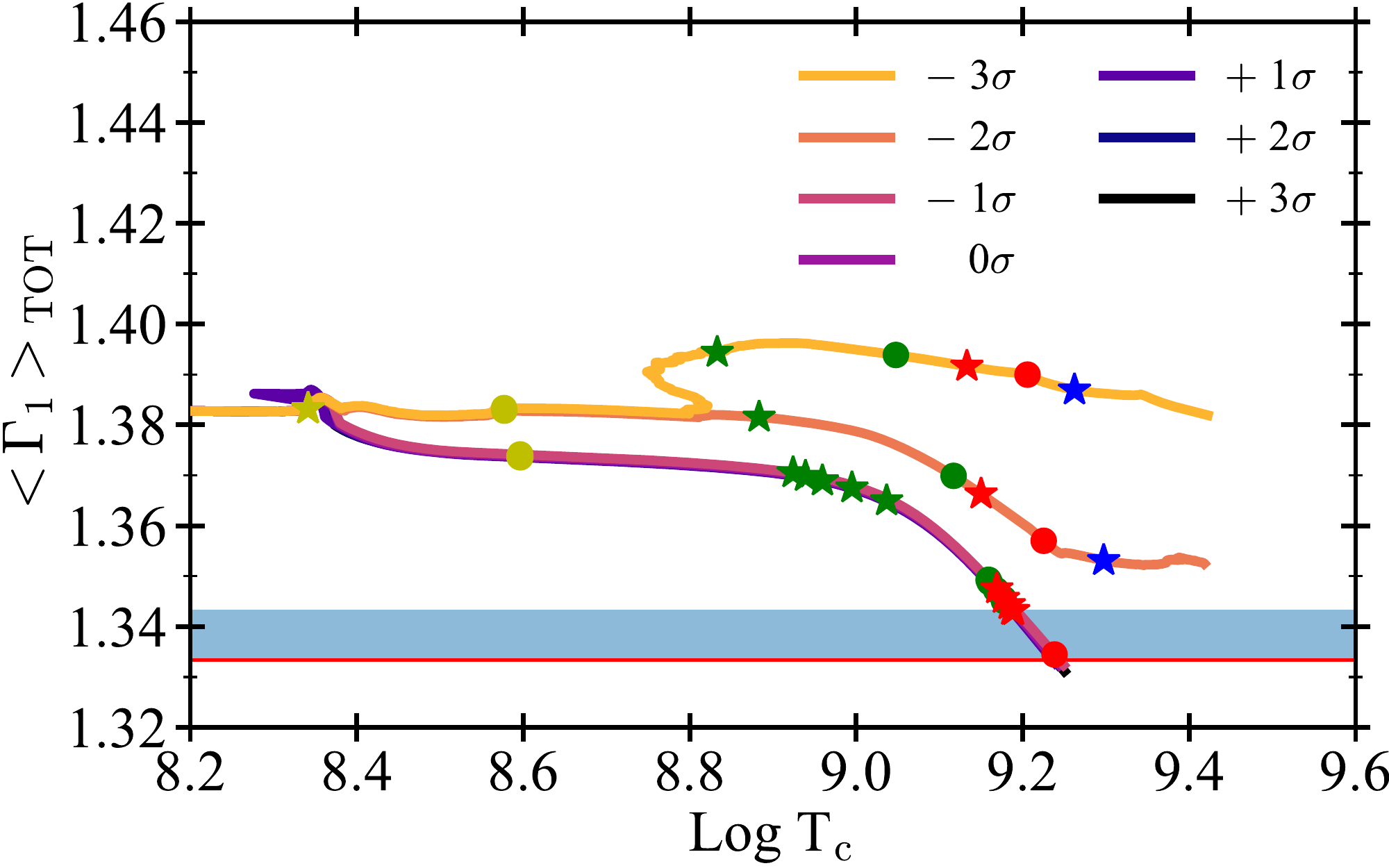} \\
            \includegraphics[width=0.48\textwidth]{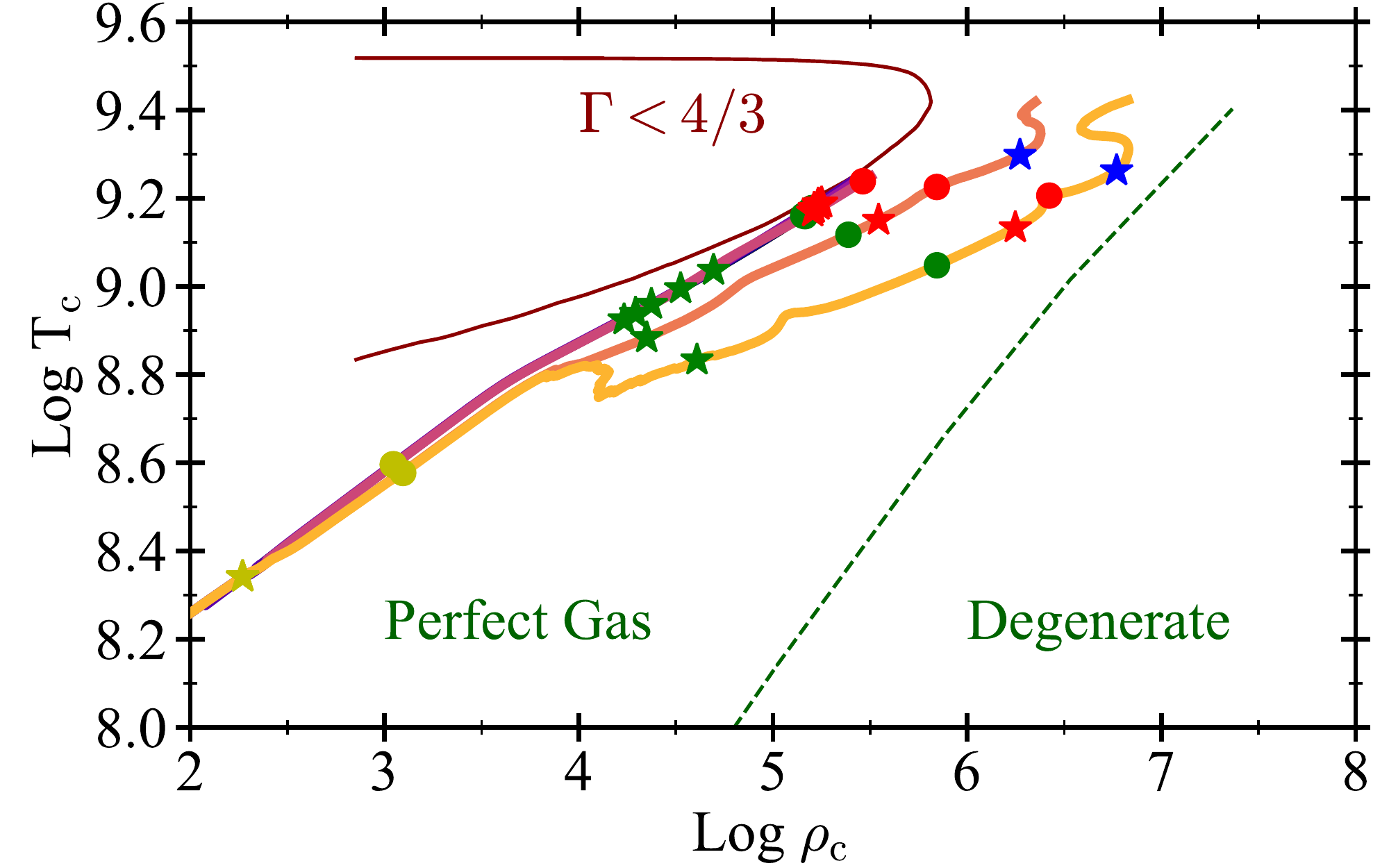}

            \caption{
            Upper panel: \G1\ values versus the central temperature of stars with M$_{\rm ZAMS}$ = 140~\Msun. 
            Lower panel: central density and temperature evolution of models with an initial mass of 140~\Msun. In both the upper and the lower right-hand panels we use the same colour code for lines and symbols as in Fig.~\ref{fig:PureHe_gamma1} and~\ref{fig:H100_gamma1}.
                    }
            \label{fig:H140_G1_rhoT}
        \end{figure}

        At this point, the star, which is well outside the PI region, evolves through the advanced burning phases until core collapse. 
        This is shown in Fig.~\ref{fig:H140_G1_rhoT} where the upper panel shows the \G1{TOT}\ value as a function of the central temperature, while the lower one shows the evolution of the central density and temperature. These plots show models with M$_{\rm ZAMS}=140$~\Msun\ computed with different \CagO\ reaction rates. The branching moment between models happens just after the beginning of the CHeB, when stars with rate $-3\, \sigma$ and $-2\, \sigma$ undergo the dredge-up, which erodes significantly the He core. The model with $-3\, \sigma$ has a second branching, when the dredge-up starts to extract matter from the CO core. The $-2\, \sigma$ model does not undergo this second branching. In all the other cases, the star evolves becoming unstable during or shortly after the neon burning phase in the core.

        {Table~\ref{tab:HTable} shows if a star undergoes the dredge-up during its evolution. It is worth to note that not all the stars that experience the dredge-up avoid the PI.}

    \subsection{Comparison with previous works}
    \label{sec:Comp_w_pre}  
    
        \citet{Takahashi2018} found that very massive stars avoid pulsational PI if they develop a carbon burning shell in the core. This happens if they assume low reaction rates.
        
        \citet{Clarkson2020} studied the interaction between convective H shell and He core in massive stars. They found that H shell and He core interact in models adopting the Schwarzschild criterion, while they do not if the Ledoux criterion is assumed.
        Models with shell interaction can produce up to 1000x more $^{14}$N in the shell with respect to models that do not have a shell interaction.

        \citet{Kaiser2020} studied the role of overshooting and semi-convection in stars with masses between 15~\Msun\ and 25~\Msun, adopting different criteria for convection. They found relative uncertainties up to about 70 \% on the core masses and lifetimes depending on the mixing assumption. Moreover, the relative importance of semi-convection decreases with an increasing amount of overshooting. The amount of overshooting is the main source of uncertainty in all phases. 
        
        Recently, \citet{Farrell2020} discuss the role of the H-shell -- He-core interaction in metal-poor massive stars. They find that uncertainties related to convective mixing, mass-loss, H-He interactions and PPI may increase the $M_{\rm gap}$ up to $\approx 85$~\Msun.

        In our case, the dredge-up is driven by envelope undershooting. If we turn off the undershooting, by assuming a $\Lambda_{\rm env}$ = 0~\Hp, the dredge-up disappears and all models with M$_{\rm ZAMS}=140$~\Msun\ become globally unstable and undergo PI (as shown in Fig.~\ref{fig:Mass_Spec}). The fact that other authors \citep[e.g.,][]{Woosley2017, Takahashi2018, Marchant2019, Farmer2020} do not observe such process in their models is due to different assumptions in the treatment of convection. For instance, stars evolved adopting the Schwarzschild criterion and without semi-convection tend to develop bigger convective regions with respect to stars computed with the Ledoux criterion, due to molecular weight gradient.

        \subsection{BH mass spectrum}
        \label{sec:MassSpec}
\begin{figure*}
            \includegraphics[width=0.48\textwidth]{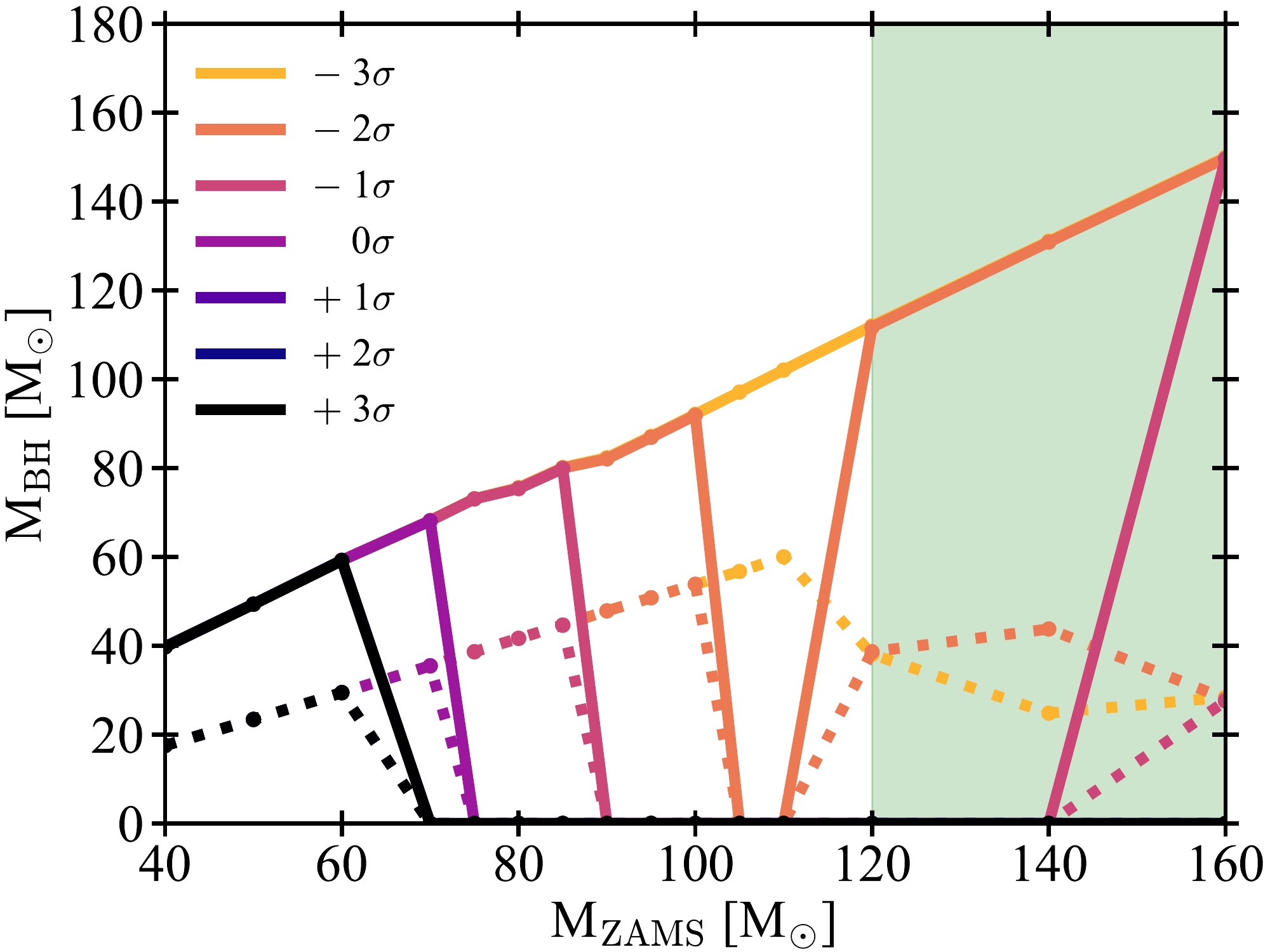}
            \includegraphics[width=0.48\textwidth]{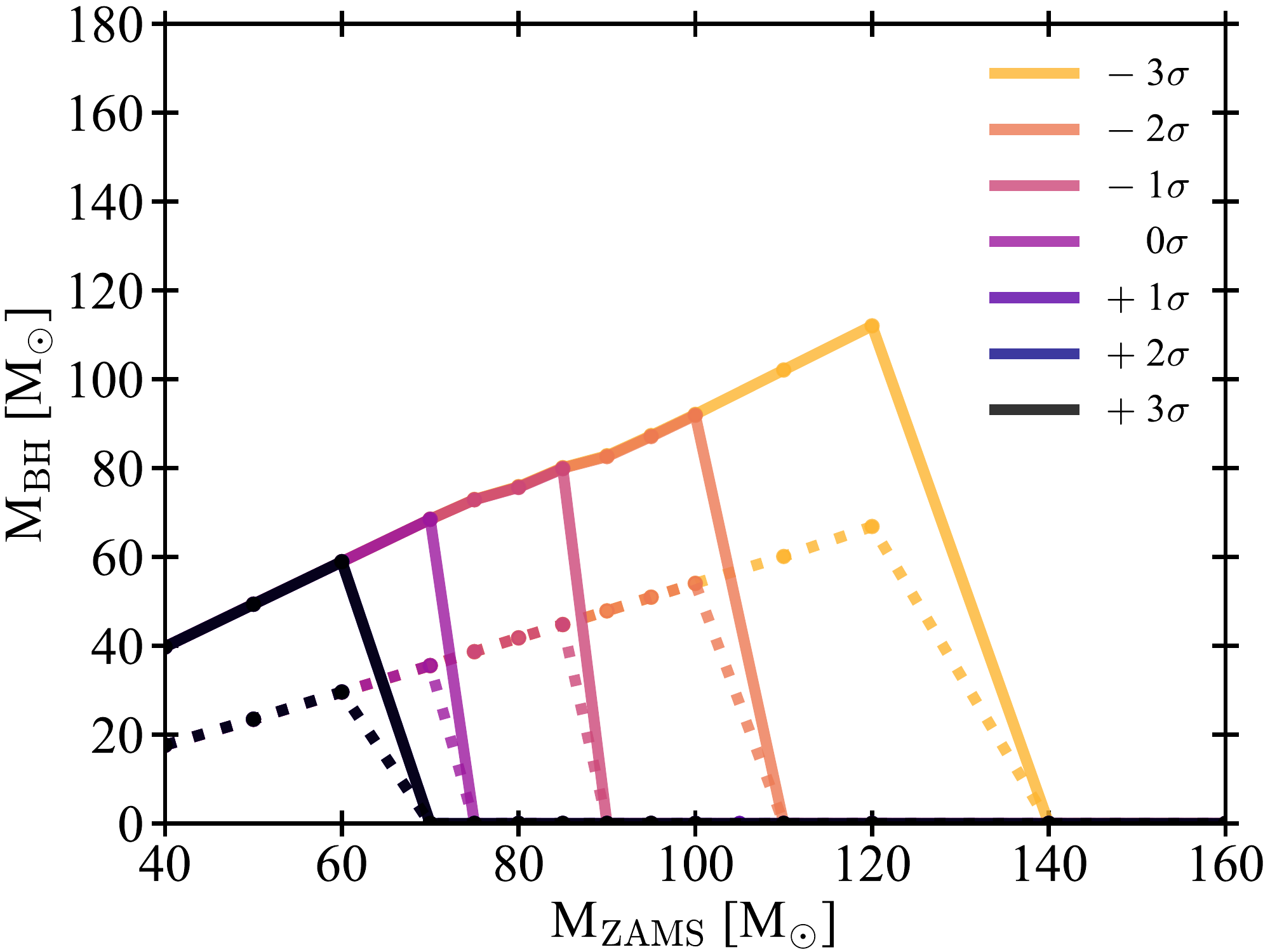}
            \caption{
            Left-hand panel: Estimated mass of the BH ($M_{\rm BH}$) as a function of the ZAMS mass of the progenitor star. Different colours indicate stars computed with varying \CagO\ reaction rates.  Solid lines indicate the total final mass of the star. Dotted lines indicate the final He core mass. As discussed in the text, the actual $M_{\rm BH}$ mass will be between the He core mass (pessimistic case, dotted lines) and the total final mass of the star (optimistic case, solid lines). The green area indicates models that experience dredge-up during the CHeB.
            Right-hand panel: Same as the left-hand panel but for a set of tracks computed without envelope undershooting.}
            \label{fig:Mass_Spec}
        \end{figure*}

            \begin{table}
                \caption{Lower edge of the PI mass gap (M$_{\rm gap}$) from stars with H-rich envelopes. }
                \begin{center} 
                \begin{tabular}{lcc} 
                    \hline\hline
                    & $\Lambda_{\rm env}$ = 0.7~\Hp & $\Lambda_{\rm env}$ = 0~\Hp \\
                    \hline
                    Rate & M$_{\rm gap}$ & M$_{\rm gap}$ \\
                     & [\Msun] & [\Msun] \\
                    \hline
                         $-3\, \sigma$ & None & 112 \\
                         $-2\, \sigma$ & 92   & 92  \\
                         $-1\, \sigma$ & 80   & 80  \\
                       ~~$0\, \sigma$  & 68   & 68  \\
                        $+1\, \sigma$  & 60   & 60  \\
                        $+2\, \sigma$  & 60   & 60  \\
                        $+3\, \sigma$  & 60   & 60  \\
                    \hline
                \end{tabular} 
                \end{center}
                \footnotesize{Column 1: rate assumed for the \CagO\ reaction; column 2: lower edge of the mass gap M$_{\rm gap}$ for models computed with envelope undershooting; column 3: M$_{\rm gap}$ for models computed without envelope undershooting. "None" indicates that no star undergoes PI.
                     }
                \label{tab:Mgap} 
            \end{table}
            The left-hand panel of Fig.~\ref{fig:Mass_Spec} shows the BH mass as a function of  M$_{\rm ZAMS}$ and for different \CagO\ reaction rates. Since we do not know what fraction of the hydrogen envelope is lost during the final collapse, we give a pessimistic and an optimistic estimate for the BH mass in Fig.~\ref{fig:Mass_Spec}, which bracket this uncertainty. In the pessimistic case (dotted line), the mass of the BH is the same as the final He core mass, which corresponds to assuming that the H envelope is completely lost during the collapse.
            In the optimistic case (solid line), the mass of the BH is the same as the entire final mass of the star, including all the residual H envelope.
            
            For stars computed with \fco\ corresponding to $+ 3\, \sigma$, $+ 2\, \sigma$ and $+ 1\, \sigma$, we find a maximum BH mass $M_{\rm BH}=60$~\Msun. For stars computed with $0 \, \sigma$, we find a maximum BH mass of 68~\Msun.
            
            The PI mass gap for models computed with $- 1\, \sigma$ is between $\approx{}80$~\Msun\ and $\approx{}150$~\Msun. In fact, the dredge-up allows stars with ZAMS mass $M_{\rm ZAMS}= 160$~\Msun\ to leave BHs with mass $\approx{}150$~\Msun.
            
            The effect of dredge-up becomes stronger if we consider lower rates for the \CagO\ reaction.
            In the $-2\, \sigma$ case, the PI mass gap is between $M_{\rm BH}=92$~\Msun\ and 110~\Msun. Finally, models computed with $- 3\, \sigma$ do not show any mass gap. The mass gap is completely removed by the effect of dredge-up.
            
            The right-hand panel of Fig.~\ref{fig:Mass_Spec} shows the same plot of the estimated BH mass but for stars evolved without envelope undershooting ($\Lambda_{\rm env}$ = 0~\Hp). In this case, stars more massive than 110~\Msun\ do not undergo the dredge-up during the CHeB phase. Without the dredge-up, the mass gap re-appears, even for low \CagO\ rates. The maximum stellar mass that ignites oxygen non-explosively is about 112~\Msun, in the case with rate $- 3\, \sigma$.
            Hence, if undershooting is suppressed, the lower edge of the mass gap $M_{\rm gap}$ is at 60~\Msun\ for rates $+3$,$+2$ and $+1$, while $M_{\rm gap} \approx 68$, 80, 92 and 112~\Msun\ for rates 0, $-1$, $-2$ and $-3\, \sigma$, respectively.
            
            Table~\ref{tab:Mgap} lists the results obtained for the lower edge of the mass gap for stars with H-rich envelopes, computed with and without envelope undershooting and adopting the optimistic case.

        %

%
%
\section{Discussion}

    The recent detection of GW190521 \citep{abbottGW190521,abbottGW190521astro}, with a primary BH mass of $85^{+21}_{-14}$ M$_{\odot}$ and a secondary BH mass of $66^{+17}_{-18}$ M$_{\odot}$ (90\% credible intervals), challenges current models of BH formation.

    Here, we have shown that the lower edge of the PI mass gap, $M_{\rm gap}$, can be as high as $\approx{}70$ M$_\odot$, if we assume that the H envelope collapses to a BH directly (Fig.~\ref{fig:Mass_Spec}). Moreover, if we not only assume that the H envelope collapses but we also take into account the uncertainty on the \CagO\ nuclear reaction rate, $M_{\rm gap}$ can be even higher. For the standard \CagO\ rate $-1\,{}\sigma$, the lower edge and the upper edge of the mass gap are $\approx{}80$ M$_{\odot}$ and $\approx{}150$ M$_{\odot}$, respectively. For the standard \CagO\ rate $-2\,{}\sigma{}$, the lower edge and the upper edge of the mass gap are $\approx{}92$ M$_{\odot}$ and $\approx{}110$ M$_{\odot}$, respectively. Finally for the standard \CagO\ rate $-3\,{}\sigma$, the PI mass gap completely disappears, because of the envelope undershooting effect we just discussed.

    The main reasons why previous work has neglected the effect of the H envelope, calculating only models of pure-He stars, is that the H envelope is usually lost during the evolution of a interacting binary star, because of Roche lobe mass transfer and common-envelope episodes. Only metal-poor single stars, or metal-poor stars in loose binary systems (with initial semi-major axis $\gtrsim{}10^3$ R$_\odot$, see e.g. Fig.~9 of \citealt{Spera2019}) can preserve most  of their H envelope to the very end of their life. 

    Loose binary systems cannot lead to BBH mergers via isolated binary evolution, because the initial orbital separation of the BBH is too large to permit coalescence by GW emission. Hence, even in our most optimistic case, it is hard to form systems like GW190521 through isolated binary evolution.

    However, in a dense stellar system, the evolution of a single BH (formed from the collapse of a single massive star) and that of a loose BBH (formed from the evolution of a loose binary star) can be very different from that of isolated binary systems  \citep[e.g.,][]{sigurdsson1995,portegieszwart2000,banerjee2010,downing2010,tanikawa2013,ziosi2014,samsing2014,morscher2015,rodriguez2016,mapelli2016,samsing2018,banerjee2017,fragionekocsis2018,arcasedda2018,banerjee2020,kumamoto2019,kumamoto2020,dicarlo2019,dicarlo2020a,dicarlo2020b,rizzuto2020,fragione2020}. The single BH can pair up with another BH via three-body encounters and dynamical exchanges. A loose BBH can harden (i.e. its semi-major axis shrinks) via three body encounters, speeding up its merger by gravitational waves. 
    The more massive a BH is, the more it is effective in acquiring companions via exchanges and in undergoing dynamical hardening \citep{heggie1975,hills1976,hills1980}. For this reason, dynamics favours the formation and merger of the most massive BBHs. Thus, the formation of a BH with mass ~85~\Msun\ from the collapse of a single star can easily lead to a system like GW190521 in a dense stellar system.  In particular, \cite{dicarlo2020b} have already shown that the dynamical formation of a merging BBH with masses similar to GW190521 (88 and 48 M$_\odot$) is possible in a young dense star cluster. 
    
    Moreover, a $\approx{}85$ M$_\odot$ BH born from the direct collapse of a star will preserve most (if not all) the spin of its progenitor star, leading to a fast rotating BH \citep{Mapelli2020}. If this BH acquires a companion via dynamical interactions, the spins of the primary and of the secondary BH will be isotropically oriented, possibly leading to a large precession spin ($\chi_{\rm p}$). This feature is consistent with the properties of GW190521 \citep{abbottGW190521,abbottGW190521astro}. 
    
    Recently, \cite{belczynski2020} even suggested that isolated binary star evolution at low metallicity can produce BBHs with masses consistent with GW190521, when the uncertainties on the \CagO\ rate are taken into account (assuming $-2.5$\,{}$\sigma{}$ rates from \citealt{Farmer2020}) and when extreme stellar masses ($M_{\rm ZAMS}\approx{}180$ M$_\odot$) are considered. However, the isolated binary evolution channel predicts a lower $\chi_{\rm p}$ than the one estimated by \cite{abbottGW190521astro} within the 90\% credible interval.

\section{Conclusions}
\label{sec:Conc}
    We investigated the lower edge of the pair-instability (PI) BH mass gap ($M_{\rm gap}$), by means of the {\sc parsec} evolutionary code.

    We evolved i) a set of pure-He stars with initial masses between 20~\Msun\ and 100~\Msun\ and with metallicity $Z=0.001$, ii) a set of stars with H envelopes, with masses ranging from 40 to 160~\Msun and with $Z=0.0003$. For both sets, we have varied the \CagO\ rates from $-3\,{}\sigma{}$ to $+3\, \sigma{}$, where $1\, \sigma$ is the given confidence error by the \textsc{starlib} library. We have followed the evolution of the models until oxygen burning, if the star remains globally stable. We find that, changing the \CagO\ rates directly affects the carbon-to-oxygen ratio at the end of the CHeB.
    
    In the case of pure-He stars, the lower edge of the mass gap ($M_{\rm gap}$) varies from 68~\Msun\ ($-3\, \sigma{}$ \CagO\ rate) to 20~\Msun\ ($+3\, \sigma$ rate). For the standard \CagO\ rate, we find $M_{\rm gap}\approx 30$~\Msun. 
    
    Since we use a hydro-static code, we do not follow the evolution through the pulsational pair-instability (PPI) phase. We conservatively assume that both stars going through PPI and stars going through pair-instability supernova (PISN) leave no compact objects. Since stars going through PPI might be able to stabilize and form BHs by direct collapse \citep{Marchant2019}, our findings are robust lower limits for the lower edge of the mass gap. 
    For this reason we find lower values of $M_{\rm gap}$ with respect to previous works \citep{Woosley2017,Leung2019,Farmer2020} in case of pure-He stars. 
    
    In the case of stellar models with an hydrogen envelope, the main uncertainty on the final BH mass is represented by the fate of the envelope \citep{Mapelli2020}. Here, we consider a pessimistic estimate in which only the He core collapses to the final BH and an optimistic approach in which the  H envelope entirely collapses to the final BH.
    
   Adopting the standard \CagO\ rate and assuming an optimistic approach, we find $M_{\rm gap}=68$~\Msun. Assuming the $+3\, \sigma$ \CagO\ rate, this value lowers to $M_{\rm gap}=59$~\Msun.

    If we consider lower values than the standard \CagO\ rate, we find that the interaction between the hydrogen envelope and the He core may result in a very different evolution of the star. Models computed with low \CagO\ rates ($-1$, $-2$, $-3\, \sigma$) and with an initial mass higher that 110~\Msun\ may experience an important dredge-up during the CHeB that extracts matter from the core enriching the envelope. 
    
    Stars experiencing such dredge-up are more stable and ignite oxygen in a non-explosive way, despite their high mass. The most remarkable result is that the PI mass gap narrows in the  $-1$ and $-2\, \sigma$ cases, and it completely disappears in the $-3\, \sigma$ case. 
    
    In the $-3\, \sigma$ case, all our models evolve until the core collapse, without undergoing PI. In case of $-2\, \sigma$ and $-1\, \sigma$ we find a mass gap between $92-110$~\Msun\ and $80-150$~\Msun, respectively.
    
    For higher rates (0, +1, +2, $+3\, \sigma{}$), the dredge-up does not happen and the lower edge of the mass gap is $M_{\rm gap}\approx{}60-70$~\Msun\ in the optimistic case ($M_{\rm gap}\approx{}30-40$~\Msun\ in the pessimistic case).
    
    We confirm that the \CagO\ reaction has a strong impact on the final fate of a massive star, as found by other authors \citep{Takahashi2018,Farmer2019,Farmer2020}.
    In addition, we show that another challenging phenomenon could be the occurrence of an efficient dredge-up at the bottom of the H-rich envelope, in the advanced phases. 
    Stars that experience such a dredge-up may develop envelopes heavily enriched by N and O. In the case such enriched envelopes are ejected, their contribution to stellar yields, especially N, may be significant. This could be relevant for the chemical evolution of nitrogen at low metallicity \citep{Pettini2008, Vincenzo2016}.

    Taking into account the uncertainties on the collapse of the H envelope and on the \CagO\ reaction rate, we can explain the formation of massive BHs such as the primary component of GW190521 \citep{abbottGW190521,abbottGW190521astro} with the collapse of a single massive metal-poor star. 
    If such massive BH forms in a dense stellar cluster, it might have a chance to dynamically pair up with another massive BH, leading to the formation of a GW190521-like system, with a large precession spin.

\section*{Acknowledgements}
   We are grateful to the anonymous referee for their useful comments. We thank Rob Farmer for the fruitful discussions and for providing us \textsc{mesa} pure-He models.
   MM, GC and GI acknowledge financial support from the European Research Council for the ERC Consolidator grant DEMOBLACK, under contract no. 770017. AB acknowledges support from PRIN-MIUR~2017 prot. 20173ML3WW. This research made use of \textsc{NumPy} \citep{Harris2020}, \textsc{SciPy} \citep{SciPy2020}, \textsc{IPython} \citep{Ipython}. For the plots we used \textsc{Matplotlib}, a Python library for publication quality graphics \citep{Hunter2007}.

\section*{Data Availability}
The data underlying this article will be shared on reasonable request to the corresponding author.



\bibliographystyle{mnras/mnras}
\bibliography{References} 

%

\appendix

\section{Tables}
\label{appendix:Tab}

Tables~\ref{tab:HeTable} and \ref{tab:HTable} show the results obtained for pure-He and full stars. More details in Sections \ref{sec:Finmass_He} and \ref{sec:Finmass_H}.
%
    \begin{table*} 
        \caption{Pure He stars' results. See text for details.} 
        \begin{center}
        \begin{tabular}{lcccccccccc} 
            \hline\hline
            M$_{\rm He-ZAMS}$ & Rate & M$_{\rm He}$ & M$_{\rm CO}$ & XC$_{\rm c-He}$ & XO$_{\rm c-He}$ & M$_{\rm Pre}$ & Log T$_{\rm c}$ & XO$_{\rm c-Pre}$ & \G1{TOT-} & Fate \\

            [\Msun] &  & [\Msun] & [\Msun] & Mass frac. & Mass frac. & [\Msun] & [K] & Mass frac. & & \\
            \hline
            20 & $-3\,\sigma$ & 19.31 & 15.36 & 0.49 & 0.49 & 19.30 & 9.378 & 0.046 & 0.0553 & CC\\
            20 & $-2\, \sigma$ & 19.29 & 15.34 & 0.41 & 0.58 & 19.28 & 9.405 & 0.013 & 0.0534 & CC\\
            20 & $-1\, \sigma$ & 19.27 & 15.28 & 0.33 & 0.66 & 19.26 & 9.419 & 0.013 & 0.0476 & CC\\
            20 & $\,\,0\, \sigma$        & 19.24 & 15.28 & 0.24 & 0.75 & 19.23 & 9.401 & 0.110 & 0.0393 & CC\\
            20 & $+1\, \sigma$ & 19.22 & 15.29 & 0.15 & 0.84 & 19.21 & 9.415 & 0.175 & 0.0256 & CC\\
            20 & $+2\, \sigma$ & 19.19 & 15.33 & 0.07 & 0.91 & 19.19 & 9.419 & 0.224 & 0.0187 & CC\\
            20 & $+3\, \sigma$ & 19.18 & 15.41 & 0.01 & 0.94 & 19.17 & 9.421 & 0.234 & 0.0175 & CC\\
            \hline
            30 & $-3\, \sigma$ & 29.07 & 24.34 & 0.46 & 0.52 & 29.09 & 9.404 & 0.075 & 0.0389  & CC\\
            30 & $-2\, \sigma$ & 29.04 & 24.29 & 0.37 & 0.61 & 29.06 & 9.409 & 0.094 & 0.0339  & CC\\
            30 & $-1\, \sigma$ & 29.01 & 24.30 & 0.29 & 0.70 & 29.03 & 9.402 & 0.129 & 0.0249  & CC\\
            30 & $\,\,0\, \sigma$        & 28.98 & 24.27 & 0.19 & 0.79 & 28.99 & 9.417 & 0.218 & 0.0096  & CC\\
            30 & $+1\, \sigma$ & 28.94 & 24.28 & 0.11 & 0.87 & 28.96 & 9.418 & 0.313 & 0.0021  & CC\\
            30 & $+2 \, \sigma$ & 28.92 & 24.31 & 0.04 & 0.92 & 28.93 & 9.417 & 0.405 & -0.0015 & PI\\
            30 & $+3\, \sigma$ & 28.90 & 24.33 & 0.01 & 0.91 & 28.92 & 9.418 & 0.387 & -0.0019 & PI\\
            \hline
            40 & $-3\, \sigma$ & 38.87 & 33.23 & 0.43 & 0.54 & 38.89 & 9.415 & 0.092 & 0.0259  & CC\\
            40 & $-2\, \sigma$ & 38.83 & 33.21 & 0.34 & 0.63 & 38.86 & 9.414 & 0.136 & 0.0185  & CC\\
            40 & $-1\, \sigma$ & 38.79 & 33.22 & 0.26 & 0.72 & 38.82 & 9.410 & 0.210 & 0.0094  & CC\\
            40 & $\,\,0\, \sigma$ & 38.76 & 33.20 & 0.17 & 0.81 & 38.78 & 9.429 & 0.270 & -0.0068 & PI\\
            40 & $+1\, \sigma$ & 38.73 & 33.22 & 0.09 & 0.88 & 38.75 & 9.326 & 0.906 & -0.0026 & PI\\
            40 & $+2\, \sigma$ & 38.69 & 33.18 & 0.03 & 0.91 & 38.72 & 9.350 & 0.923 & -0.0063 & PI\\
            40 & $+3\, \sigma$ & 38.67 & 33.19 & 0.00 & 0.89 & 38.70 & 9.397 & 0.698 & -0.0098 & PI\\
            \hline
            50 & $-3\, \sigma$ & 48.68 & 42.16 & 0.41 & 0.56 & 48.72 & 9.404 & 0.098 & 0.0305  & CC\\
            50 & $-2\, \sigma$ & 48.64 & 42.16 & 0.32 & 0.65 & 48.68 & 9.396 & 0.126 & 0.0247  & CC\\
            50 & $-1\, \sigma$ & 48.60 & 42.08 & 0.23 & 0.73 & 48.63 & 9.411 & 0.306 & -0.0047 & PI\\
            50 & $\,\,0\, \sigma$        & 48.56 & 42.12 & 0.15 & 0.82 & 48.59 & 9.379 & 0.719 & -0.0108 & PI\\
            50 & $+1\, \sigma$ & 48.51 & 42.12 & 0.07 & 0.88 & 48.55 & 9.294 & 0.915 & -0.0069 & PI\\
            50 & $+2\, \sigma$ & 48.48 & 42.11 & 0.02 & 0.90 & 48.52 & 9.269 & 0.933 & -0.0030 & PI\\
            50 & $+3\, \sigma$ & 48.46 & 42.04 & 0.00 & 0.87 & 48.49 & 9.284 & 0.916 & -0.0057 & PI\\
            \hline
            55 & $-3\, \sigma$ & 53.60 & 46.68 & 0.40 & 0.56 & 53.64 & 9.406 & 0.107 & 0.0276  & CC\\
            55 & $-2\, \sigma$ & 53.55 & 46.56 & 0.31 & 0.65 & 53.59 & 9.414 & 0.138 & 0.0202  & CC\\
            55 & $-1\, \sigma$ & 53.51 & 46.62 & 0.22 & 0.74 & 53.55 & 9.397 & 0.617 & -0.0088 & PI\\
            55 & $\,\,0\, \sigma$        & 53.46 & 46.57 & 0.14 & 0.82 & 53.50 & 9.287 & 0.870 & -0.0090 & PI\\
            55 & $+1\, \sigma$ & 53.42 & 46.53 & 0.06 & 0.88 & 53.46 & 9.276 & 0.916 & -0.0077 & PI\\
            55 & $+2\, \sigma$ & 53.39 & 46.57 & 0.02 & 0.89 & 53.42 & 9.244 & 0.917 & -0.0019 & PI\\
            55 & $+3\, \sigma$ & 53.36 & 46.53 & 0.00 & 0.86 & 53.40 & 9.265 & 0.909 & -0.0059 & PI\\
            \hline
            60 & $-3\, \sigma$ & 58.51 & 51.20 & 0.39 & 0.57 & 58.56 & 9.399 & 0.102 & 0.0237  & CC\\
            60 & $-2\, \sigma$ & 58.47 & 51.08 & 0.30 & 0.66 & 58.51 & 9.381 & 0.147 & 0.0162  & CC\\
            60 & $-1\, \sigma$ & 58.43 & 51.06 & 0.22 & 0.74 & 58.46 & 9.358 & 0.707 & -0.0108 & PI\\
            60 & $\,\,0\, \sigma$        & 58.37 & 51.04 & 0.13 & 0.82 & 58.41 & 9.272 & 0.872 & -0.0098 & PI\\
            60 & $+1\, \sigma$ & 58.33 & 51.02 & 0.06 & 0.88 & 58.37 & 9.266 & 0.914 & -0.0093 & PI\\
            60 & $+2\, \sigma$ & 58.29 & 51.03 & 0.02 & 0.88 & 58.33 & 9.250 & 0.914 & -0.0063 & PI\\
            60 & $+3\, \sigma$ & 58.27 & 51.08 & 0.00 & 0.85 & 58.31 & 9.263 & 0.901 & -0.0089 & PI\\
            \hline
            65 & $-3\, \sigma$ & 63.44 & 55.61 & 0.38 & 0.57 & 63.48 & 9.409 & 0.126 & 0.0215  & CC\\
            65 & $-2\, \sigma$ & 63.38 & 55.64 & 0.30 & 0.66 & 63.43 & 9.381 & 0.166 & 0.0142  & CC\\
            65 & $-1\, \sigma$ & 63.33 & 55.63 & 0.21 & 0.75 & 63.38 & 9.275 & 0.815 & -0.0098 & PI\\
            65 & $\,\,0\, \sigma$        & 63.29 & 55.56 & 0.12 & 0.82 & 63.33 & 9.258 & 0.873 & -0.0102 & PI\\
            65 & $+1\, \sigma$ & 63.24 & 55.55 & 0.05 & 0.87 & 63.28 & 9.251 & 0.905 & -0.0093 & PI\\
            65 & $+2\, \sigma$ & 63.20 & 55.49 & 0.01 & 0.88 & 63.25 & 9.250 & 0.907 & -0.0091 & PI\\
            65 & $+3\, \sigma$ & 63.18 & 55.44 & 0.00 & 0.85 & 63.22 & 9.249 & 0.885 & -0.0090 & PI\\
            \hline
            70 & $-3\, \sigma$ & 68.35 & 60.15 & 0.38 & 0.58 & 68.41 & 9.418 & 0.105 & 0.0186  & CC\\
            70 & $-2\, \sigma$ & 68.30 & 60.13 & 0.29 & 0.67 & 68.35 & 9.375 & 0.653 & -0.0099 & PI\\
            70 & $-1\, \sigma$ & 68.25 & 60.15 & 0.20 & 0.75 & 68.30 & 9.261 & 0.817 & -0.0111 & PI\\
            70 & $\,\,0\, \sigma$        & 68.20 & 60.00 & 0.12 & 0.83 & 68.25 & 9.248 & 0.869 & -0.0110 & PI\\
            70 & $+1\, \sigma$ & 68.16 & 60.02 & 0.05 & 0.87 & 68.20 & 9.244 & 0.896 & -0.0104 & PI\\
            70 & $+2\, \sigma$ & 68.12 & 60.02 & 0.01 & 0.87 & 68.16 & 9.234 & 0.885 & -0.0085 & PI\\
            70 & $+3\, \sigma$ & 68.09 & 59.96 & 0.00 & 0.84 & 68.14 & 9.238 & 0.865 & -0.0094 & PI\\
            \hline 
        \end{tabular} 
        \end{center}
        \label{tab:HeTable} 
    \end{table*}
    \begin{table*} 
        \contcaption{} 
        \begin{center}
        \begin{tabular}{lcccccccccc} 
            \hline\hline
            M$_{\rm He-ZAMS}$ & Rate & M$_{\rm He}$ & M$_{\rm CO}$ & XC$_{\rm c-He}$ & XO$_{\rm c-He}$ & M$_{\rm Pre}$ & Log T$_{\rm c}$ & XO$_{\rm c-Pre}$ & \G1{TOT-} & Fate \\

            [\Msun] &  & [\Msun] & [\Msun] & Mass frac. & Mass frac. & [\Msun] & [K] & Mass frac. & & \\

                    \hline
            80 & $-3\, \sigma$ & 78.21 & 69.19 & 0.37 & 0.58 & 78.26 & 9.316 & 0.681 & -0.0076 & PI\\
            80 & $-2\, \sigma$ & 78.15 & 69.17 & 0.28 & 0.67 & 78.21 & 9.247 & 0.759 & -0.0095 & PI\\
            80 & $-1\, \sigma$ & 78.09 & 69.16 & 0.19 & 0.75 & 78.15 & 9.227 & 0.795 & -0.0109 & PI\\
            80 & $\,\,0\, \sigma$        & 78.02 & 69.07 & 0.11 & 0.83 & 78.09 & 9.226 & 0.836 & -0.0113 & PI\\
            80 & $+1\, \sigma$ & 77.98 & 69.00 & 0.04 & 0.87 & 78.04 & 9.220 & 0.863 & -0.0102 & PI\\
            80 & $+2\, \sigma$ & 77.94 & 68.96 & 0.01 & 0.86 & 78.00 & 9.221 & 0.863 & -0.0105 & PI\\
            80 & $+3\, \sigma$ & 77.91 & 68.91 & 0.00 & 0.83 & 77.98 & 9.222 & 0.837 & -0.0107 & PI\\
            \hline
            90 & $-3\, \sigma$ & 88.06 & 78.15 & 0.36 & 0.59 & 88.13 & 9.232 & 0.695 & -0.0087 & PI\\
            90 & $-2\, \sigma$ & 87.99 & 78.05 & 0.27 & 0.68 & 88.06 & 9.203 & 0.666 & -0.0095 & PI\\
            90 & $-1\, \sigma$ & 87.94 & 78.13 & 0.18 & 0.76 & 88.00 & 9.204 & 0.737 & -0.0103 & PI\\
            90 & $\,\,0\, \sigma$        & 87.87 & 78.05 & 0.10 & 0.83 & 87.94 & 9.201 & 0.804 & -0.0100 & PI\\
            90 & $+1\, \sigma$ & 87.82 & 78.00 & 0.04 & 0.86 & 87.89 & 9.204 & 0.851 & -0.0107 & PI\\
            90 & $+2\, \sigma$ & 87.78 & 77.97 & 0.01 & 0.85 & 87.85 & 9.201 & 0.848 & -0.0099 & PI\\
            90 & $+3\, \sigma$ & 87.75 & 77.96 & 0.00 & 0.82 & 87.82 & 9.202 & 0.821 & -0.0103 & PI\\
            \hline
            100 & $-3\, \sigma$ & 97.92 & 87.20 & 0.35 & 0.59 & 97.99 & 9.194 & 0.562 & -0.0105 & PI\\
            100 & $-2\, \sigma$ & 97.84 & 87.10 & 0.26 & 0.68 & 97.92 & 9.193 & 0.642 & -0.0110 & PI\\
            100 & $-1\, \sigma$ & 97.78 & 87.08 & 0.17 & 0.76 & 97.86 & 9.194 & 0.726 & -0.0115 & PI\\
            100 & $\,\,0\, \sigma$        & 97.71 & 87.08 & 0.09 & 0.82 & 97.79 & 9.192 & 0.801 & -0.0111 & PI\\
            100 & $+1\, \sigma$ & 97.66 & 87.03 & 0.04 & 0.85 & 97.74 & 9.190 & 0.845 & -0.0107 & PI\\
            100 & $+2\, \sigma$ & 97.61 & 87.03 & 0.01 & 0.84 & 97.69 & 9.188 & 0.839 & -0.0102 & PI\\
            100 & $+3\, \sigma$ & 97.59 & 86.96 & 0.00 & 0.81 & 97.67 & 9.190 & 0.811 & -0.0107 & PI\\
                    \hline 
        \end{tabular}
        \end{center}
        \footnotesize{Column 1: He-ZAMS mass. Column 2: rate assumed for the \CagO\ reaction. Column 3: mass of the helium core at the end of the He-MS. Column 4: mass of the CO core at the end of the He-MS. Columns 5 and 6: carbon and oxygen central mass fractions at the end of the He-MS, respectively. Column 7: final pre-supernova mass (M$_{\rm Pre}$). Column 8: central temperature of the last model. Column 9: oxygen central mass fraction at the end of the computation. Column 10: \G1{TOT-}. Column 11: final fate of the star, which can be either core collapse (CC) or pair-instability (PI). 

        }
        \label{tab:HeTable2} 
    \end{table*}
    %
    \begin{table*} 
        \caption{Results for stars with hydrogen envelopes. See text for details.}
        \centering 
        \begin{tabular}{lcccccccccccc} 
            \hline\hline
            M$_{\rm ZAMS}$ & Rate & M$_{\rm He}$ & M$_{\rm CO}$ & XC$_\mathrm{c-He}$ & XO$_\mathrm{c-He}$ & M$_\mathrm{Pre}$ & Log T$_\mathrm{c}$ & XO$_\mathrm{c-Pre}$ & \G1{Core}$_-$ & \G1{TOT}$_-$ & Fate & Dredge-up\\

            [\Msun] &  & [\Msun] & [\Msun] & Mass frac. & Mass frac. & [\Msun] & [K] & Mass frac. & & \\
            \hline\\

                40 & $-3\, \sigma$ & 17.50 & 13.81 & 0.50 & 0.49 & 39.75 & 9.365 & 0.013 & 0.0617 & 0.0617    & CC & No\\
                40 & $-2\, \sigma$ & 17.51 & 13.81 & 0.42 & 0.57 & 39.75 & 9.185 & 0.627 & 0.0603 & 0.0604    & CC & No\\
                40 & $-1\, \sigma$ & 17.51 & 13.81 & 0.34 & 0.66 & 39.76 & 9.409 & 0.011 & 0.0544 & 0.0546    & CC & No\\
                40 & $ 0 \sigma$ & 17.51 & 13.81 & 0.24 & 0.75 & 39.76 & 9.397 & 0.087 & 0.0470 & 0.0474    & CC & No\\
                40 & $+1\, \sigma$ & 17.52 & 13.82 & 0.15 & 0.84 & 39.76 & 9.400 & 0.257 & 0.0361 & 0.0367    & CC & No\\
                40 & $+2\, \sigma$ & 17.52 & 13.82 & 0.07 & 0.91 & 39.76 & 9.378 & 0.522 & 0.0334 & 0.0341    & CC & No\\
                40 & $+3\, \sigma$ & 17.53 & 13.82 & 0.01 & 0.94 & 39.76 & 9.417 & 0.205 & 0.0272 & 0.0279    & CC & No\\
            \hline
                50 & $-3\, \sigma$ & 23.42 & 19.15 & 0.47 & 0.51 & 49.39 & 9.189 & 0.582 & 0.0489 & 0.0488    & CC & No\\
                50 & $-2\, \sigma$ & 23.42 & 19.16 & 0.39 & 0.60 & 49.39 & 9.402 & 0.070 & 0.0450 & 0.0451    & CC & No\\
                50 & $-1\, \sigma$ & 23.42 & 19.16 & 0.30 & 0.68 & 49.39 & 9.440 & 0.000 & 0.0364 & 0.0366    & CC & No\\
                50 & $\,\,0\, \sigma$  & 23.42 & 19.16 & 0.21 & 0.77 & 49.38 & 9.411 & 0.144 & 0.0294 & 0.0297    & CC & No\\
                50 & $+1\, \sigma$ & 23.42 & 19.16 & 0.12 & 0.86 & 49.37 & 9.400 & 0.323 & 0.0216 & 0.0220    & CC & No\\
                50 & $+2\, \sigma$ & 23.42 & 19.16 & 0.05 & 0.92 & 49.36 & 9.421 & 0.239 & 0.0129 & 0.0134    & CC & No\\
                50 & $+3\, \sigma$ & 23.42 & 19.17 & 0.01 & 0.93 & 49.35 & 9.422 & 0.242 & 0.0120 & 0.0126    & CC & No\\
            \hline
                60 & $-3\, \sigma$ & 29.42 & 24.63 & 0.45 & 0.53 & 59.24 & 9.405 & 0.077 & 0.0383 & 0.0384    & CC & No\\
                60 & $-2\, \sigma$ & 29.42 & 24.64 & 0.37 & 0.62 & 59.24 & 9.435 & 0.016 & 0.0323 & 0.0326    & CC & No\\
                60 & $-1\, \sigma$ & 29.42 & 24.62 & 0.28 & 0.70 & 59.24 & 9.388 & 0.209 & 0.0264 & 0.0268    & CC & No\\
                60 & $\,\,0\, \sigma$  & 29.45 & 24.63 & 0.19 & 0.79 & 59.22 & 9.416 & 0.181 & 0.0162 & 0.0168    & CC & No\\
                60 & $+1\, \sigma$ & 29.45 & 24.63 & 0.10 & 0.87 & 59.21 & 9.422 & 0.237 & 0.0057 & 0.0064    & CC & No\\
                60 & $+2\, \sigma$ & 29.45 & 24.63 & 0.04 & 0.92 & 59.20 & 9.421 & 0.312 & 0.0018 & 0.0026    & CC & No\\
                60 & $+3\, \sigma$ & 29.45 & 24.63 & 0.01 & 0.91 & 59.18 & 9.419 & 0.332 & 0.0016 & 0.0024    & CC & No\\
            \hline 
        \end{tabular} 
        \label{tab:HTable} 
    \end{table*}
    \begin{table*} 
        \contcaption{} 
        \centering 
        \begin{tabular}{lcccccccccccc} 
            \hline\hline
            M$_{\rm ZAMS}$ & Rate & M$_{\rm He}$ & M$_{\rm CO}$ & XC$_\mathrm{c-He}$ & XO$_\mathrm{c-He}$ & M$_\mathrm{Pre}$ & Log T$_\mathrm{c}$ & XO$_\mathrm{c-Pre}$ & \G1{Core}$_-$ & \G1{TOT}$_-$ & Fate & Dredge-up \\

            [\Msun] &  & [\Msun] & [\Msun] & Mass frac. & Mass frac. & [\Msun] & [K] & Mass frac. & & \\
            \hline
                70 & $-3\, \sigma$ & 35.42 & 30.25 & 0.43 & 0.54 & 68.21 & 9.441 & 0.001 & 0.0278 & 0.0280    & CC & No\\
                70 & $-2\, \sigma$ & 35.42 & 30.25 & 0.35 & 0.63 & 68.17 & 9.438 & 0.033 & 0.0229 & 0.0232    & CC & No\\
                70 & $-1\, \sigma$ & 35.42 & 30.25 & 0.26 & 0.72 & 68.15 & 9.417 & 0.160 & 0.0151 & 0.0156    & CC & No\\
                70 & $\,\,0\, \sigma$  & 35.42 & 30.38 & 0.17 & 0.80 & 68.09 & 9.420 & 0.243 & 0.0028 & 0.0035    & CC & No\\
                70 & $+1\, \sigma$ & 35.42 & 30.38 & 0.09 & 0.88 & 68.08 & 9.418 & 0.351 & -0.0037 & -0.0030  & PI & No\\
                70 & $+2\, \sigma$ & 35.46 & 30.37 & 0.03 & 0.91 & 68.07 & 9.427 & 0.302 & -0.0076 & -0.0068  & PI & No\\
                70 & $+3\, \sigma$ & 35.46 & 30.38 & 0.00 & 0.89 & 68.09 & 9.414 & 0.458 & -0.0065 & -0.0057  & PI & No\\
            \hline
                75 & $-3\, \sigma$ & 38.58 & 33.32 & 0.43 & 0.55 & 73.10 & 9.413 & 0.103 & 0.0255 & 0.0258    & CC & No\\
                75 & $-2\, \sigma$ & 38.58 & 33.32 & 0.34 & 0.64 & 73.07 & 9.415 & 0.130 & 0.0184 & 0.0188    & CC & No\\
                75 & $-1\, \sigma$ & 38.58 & 33.32 & 0.25 & 0.72 & 73.05 & 9.394 & 0.330 & 0.0116 & 0.0121    & CC & No\\
                75 & $\,\,0\, \sigma$  & 38.58 & 33.32 & 0.16 & 0.81 & 73.03 & 9.417 & 0.305 & -0.0021 & -0.0013  & PI & No\\
                75 & $+1\, \sigma$ & 38.58 & 33.33 & 0.08 & 0.88 & 73.01 & 9.400 & 0.581 & -0.0055 & -0.0047  & PI & No\\
                75 & $+2\, \sigma$ & 38.62 & 33.33 & 0.02 & 0.91 & 73.04 & 9.422 & 0.410 & -0.0108 & -0.0099  & PI & No\\
                75 & $+3\, \sigma$ & 38.62 & 33.33 & 0.00 & 0.89 & 73.05 & 9.402 & 0.643 & -0.0085 & -0.0076  & PI & No\\
            \hline
                80 & $-3\, \sigma$ & 41.54 & 35.96 & 0.42 & 0.55 & 75.60 & 9.446 & 0.001 & 0.0195 & 0.0197    & CC & No\\
                80 & $-2\, \sigma$ & 41.54 & 35.96 & 0.33 & 0.64 & 75.48 & 9.417 & 0.142 & 0.0155 & 0.0158    & CC & No\\
                80 & $-1\, \sigma$ & 41.54 & 35.95 & 0.25 & 0.73 & 75.37 & 9.420 & 0.187 & 0.0056 & 0.0061    & CC & No\\
                80 & $\,\,0\, \sigma$  & 41.54 & 35.98 & 0.16 & 0.81 & 75.25 & 9.400 & 0.495 & -0.0022 & -0.0015  & PI & No\\
                80 & $+1\, \sigma$ & 41.54 & 35.97 & 0.08 & 0.88 & 75.15 & 9.400 & 0.617 & -0.0091 & -0.0083  & PI & No\\
                80 & $+2\, \sigma$ & 41.59 & 36.13 & 0.02 & 0.90 & 75.07 & 9.418 & 0.458 & -0.0135 & -0.0127  & PI & No\\
                80 & $+3\, \sigma$ & 41.59 & 36.14 & 0.00 & 0.88 & 75.05 & 9.422 & 0.373 & -0.0140 & -0.0132  & PI & No\\
            \hline
                85 & $-3\, \sigma$ & 44.51 & 38.80 & 0.41 & 0.56 & 80.13 & 9.415 & 0.117 & 0.0192 & 0.0195    & CC & No\\
                85 & $-2\, \sigma$ & 44.51 & 38.81 & 0.33 & 0.65 & 80.00 & 9.417 & 0.148 & 0.0122 & 0.0126    & CC & No\\
                85 & $-1\, \sigma$ & 44.51 & 38.81 & 0.24 & 0.73 & 79.88 & 9.421 & 0.201 & 0.0013 & 0.0019    & CC & No\\
                85 & $\,\,0\, \sigma$  & 44.51 & 38.81 & 0.15 & 0.82 & 79.77 & 9.414 & 0.372 & -0.0083 & -0.0075  & PI & No\\
                85 & $+1\, \sigma$ & 44.51 & 38.81 & 0.07 & 0.88 & 79.67 & 9.400 & 0.647 & -0.0126 & -0.0118  & PI & No\\
                85 & $+2\, \sigma$ & 44.51 & 38.81 & 0.02 & 0.90 & 79.44 & 9.290 & 0.936 & -0.0015 & -0.0007  & PI & No\\
                85 & $+3\, \sigma$ & 44.51 & 38.81 & 0.00 & 0.87 & 79.57 & 9.404 & 0.648 & -0.0147 & -0.0138  & PI & No\\
            \hline
                90 & $-3\, \sigma$ & 47.68 & 41.70 & 0.41 & 0.56 & 82.34 & 9.404 & 0.101 & 0.0151 & 0.0154    & CC & No\\
                90 & $-2\, \sigma$ & 47.71 & 41.96 & 0.32 & 0.65 & 82.07 & 9.419 & 0.160 & 0.0070 & 0.0075    & CC & No\\
                90 & $-1\, \sigma$ & 47.71 & 41.95 & 0.23 & 0.74 & 81.85 & 9.408 & 0.331 & -0.0024 & -0.0017  & PI & No\\
                90 & $\,\,0\, \sigma$  & 47.65 & 41.96 & 0.14 & 0.82 & 81.84 & 9.404 & 0.518 & -0.0108 & -0.0100  & PI & No\\
                90 & $+1\, \sigma$ & 47.71 & 41.96 & 0.07 & 0.88 & 81.42 & 9.383 & 0.750 & -0.0147 & -0.0137  & PI & No\\
                90 & $+2\, \sigma$ & 47.71 & 41.71 & 0.02 & 0.90 & 81.31 & 9.367 & 0.895 & -0.0142 & -0.0133  & PI & No\\
                90 & $+3\, \sigma$ & 47.71 & 41.96 & 0.00 & 0.87 & 81.17 & 9.356 & 0.902 & -0.0135 & -0.0126  & PI & No\\
            \hline 
                95 & $-3\, \sigma$ & 50.59 & 44.71 & 0.40 & 0.56 & 87.14 & 9.418 & 0.130 & 0.0131 & 0.0134    & CC & No\\
                95 & $-2\, \sigma$ & 50.59 & 44.71 & 0.31 & 0.65 & 86.92 & 9.421 & 0.167 & 0.0035 & 0.0040    & CC & No\\
                95 & $-1\, \sigma$ & 50.59 & 44.72 & 0.23 & 0.74 & 86.70 & 9.405 & 0.397 & -0.0054 & -0.0047  & PI & No\\
                95 & $\,\,0\, \sigma$  & 50.59 & 44.72 & 0.14 & 0.82 & 86.46 & 9.353 & 0.776 & -0.0101 & -0.0093  & PI & No\\
                95 & $+1\, \sigma$ & 50.59 & 44.72 & 0.06 & 0.88 & 86.25 & 9.340 & 0.909 & -0.0136 & -0.0127  & PI & No\\
                95 & $+2\, \sigma$ & 50.59 & 44.72 & 0.02 & 0.89 & 86.09 & 9.332 & 0.928 & -0.0130 & -0.0121  & PI & No\\
                95 & $+3\, \sigma$ & 50.59 & 44.72 & 0.00 & 0.86 & 85.98 & 9.332 & 0.910 & -0.0130 & -0.0122  & PI & No\\
            \hline
                100 & $-3\, \sigma$ & 53.61 & 47.60 & 0.40 & 0.57 & 92.10 & 9.418 & 0.138 & 0.0100 & 0.0104   & CC & No\\
                100 & $-2\, \sigma$ & 53.61 & 47.61 & 0.31 & 0.66 & 91.88 & 9.420 & 0.175 & 0.0017 & 0.0023   & CC & No\\
                100 & $-1\, \sigma$ & 53.61 & 47.32 & 0.22 & 0.74 & 91.70 & 9.402 & 0.444 & -0.0076 & -0.0068 & PI & No\\
                100 & $\,\,0\, \sigma$  & 53.64 & 47.61 & 0.13 & 0.82 & 91.43 & 9.329 & 0.854 & -0.0114 & -0.0106 & PI & No\\
                100 & $+1\, \sigma$ & 53.64 & 47.61 & 0.06 & 0.88 & 91.22 & 9.314 & 0.916 & -0.0126 & -0.0117 & PI & No\\
                100 & $+2\, \sigma$ & 53.64 & 47.61 & 0.02 & 0.89 & 91.05 & 9.316 & 0.927 & -0.0132 & -0.0123 & PI & No\\
                100 & $+3\, \sigma$ & 53.64 & 47.61 & 0.00 & 0.86 & 90.97 & 9.316 & 0.908 & -0.0132 & -0.0123 & PI & No\\
            \hline
                105 & $-3\, \sigma$ & 56.53 & 50.04 & 0.39 & 0.57 & 97.10 & 9.420 & 0.137 & 0.0084 & 0.0088   & CC & No\\
                105 & $-2\, \sigma$ & 56.53 & 50.04 & 0.30 & 0.66 & 96.89 & 9.421 & 0.180 & -0.0006 & -0.0000 & PI & No\\
                105 & $-1\, \sigma$ & 56.53 & 50.04 & 0.22 & 0.74 & 96.67 & 9.397 & 0.521 & -0.0106 & -0.0098 & PI & No\\
                105 & $\,\,0\, \sigma$  & 56.53 & 50.05 & 0.13 & 0.82 & 96.44 & 9.326 & 0.861 & -0.0134 & -0.0125 & PI & No\\
                105 & $+1\, \sigma$ & 56.58 & 50.04 & 0.06 & 0.88 & 96.24 & 9.300 & 0.916 & -0.0123 & -0.0115 & PI & No\\
                105 & $+2\, \sigma$ & 56.58 & 50.37 & 0.01 & 0.88 & 96.03 & 9.298 & 0.925 & -0.0125 & -0.0116 & PI & No\\
                105 & $+3\, \sigma$ & 56.58 & 50.04 & 0.00 & 0.85 & 95.98 & 9.299 & 0.906 & -0.0125 & -0.0116 & PI & No\\
            \hline 
        \end{tabular} 
        \label{tab:HTable2} 
    \end{table*}
    \begin{table*} 
        \contcaption{} 
        \centering 
        \begin{tabular}{lcccccccccccc} 
            \hline\hline
            M$_{\rm ZAMS}$ & Rate & M$_{\rm He}$ & M$_{\rm CO}$ & XC$_\mathrm{c-He}$ & XO$_\mathrm{c-He}$ & M$_\mathrm{Pre}$ & Log T$_\mathrm{c}$ & XO$_\mathrm{c-Pre}$ & \G1{Core}$_-$ & \G1{TOT}$_-$ & Fate & Dredge-up \\

            [\Msun] &  & [\Msun] & [\Msun] & Mass frac. & Mass frac. & [\Msun] & [K] & Mass frac. & & \\
            \hline
                110 & $-3\, \sigma$ & 59.82 & 53.19 & 0.39 & 0.57 & 102.06 & 9.421 & 0.145 & 0.0052 & 0.0057   & CC & No\\
                110 & $-2\, \sigma$ & 59.82 & 53.19 & 0.30 & 0.66 & 101.83 & 9.421 & 0.209 & -0.0063 & -0.0056 & PI & No\\
                110 & $-1\, \sigma$ & 59.82 & 53.20 & 0.21 & 0.75 & 101.62 & 9.385 & 0.644 & -0.0127 & -0.0118 & PI & No\\
                110 & $\,\,0\, \sigma$  & 59.82 & 53.21 & 0.12 & 0.82 & 101.39 & 9.290 & 0.875 & -0.0123 & -0.0114 & PI & No\\
                110 & $+1\, \sigma$ & 59.82 & 53.21 & 0.06 & 0.88 & 101.19 & 9.287 & 0.916 & -0.0124 & -0.0115 & PI & No\\
                110 & $+2\, \sigma$ & 59.82 & 53.21 & 0.01 & 0.88 & 101.05 & 9.287 & 0.922 & -0.0127 & -0.0118 & PI & No\\
                110 & $+3\, \sigma$ & 59.88 & 53.20 & 0.00 & 0.85 & 100.95 & 9.289 & 0.902 & -0.0131 & -0.0121 & PI & No\\
            \hline
                120 & $-3\, \sigma$ & 41.73 & 37.85 & 0.40 & 0.57 & 112.00 & 9.414 & 0.110 & 0.0228 & 0.0228   & CC & Yes\\
                120 & $-2\, \sigma$ & 42.37 & 37.86 & 0.31 & 0.66 & 111.69 & 9.404 & 0.224 & 0.0147 & 0.0146   & CC & Yes\\
                120 & $-1\, \sigma$ & 61.08 & 53.34 & 0.21 & 0.75 & 115.25 & 9.251 & 0.817 & -0.0062 & -0.0054 & PI & No\\
                120 & $\,\,0\, \sigma$  & 42.96 & 38.08 & 0.14 & 0.82 & 111.26 & 9.410 & 0.384 & -0.0057 & -0.0057 & PI & Yes\\
                120 & $+1\, \sigma$ & 42.87 & 37.79 & 0.06 & 0.88 & 111.13 & 9.400 & 0.610 & -0.0093 & -0.0092 & PI & Yes\\
                120 & $+2\, \sigma$ & 60.69 & 53.18 & 0.01 & 0.88 & 115.18 & 9.280 & 0.921 & -0.0126 & -0.0116 & PI & No\\
                120 & $+3\, \sigma$ & 60.81 & 53.36 & 0.00 & 0.85 & 115.13 & 9.280 & 0.900 & -0.0126 & -0.0117 & PI & No\\
            \hline
                140 & $-3\, \sigma$ & 48.93 & 45.50 & 0.39 & 0.58 & 131.07 & 9.422 & 0.015 & 0.0389 & 0.0386   & CC & Yes\\
                140 & $-2\, \sigma$ & 49.62 & 45.37 & 0.30 & 0.67 & 130.87 & 9.418 & 0.161 & 0.0092 & 0.0092   & CC & Yes\\
                140 & $-1\, \sigma$ & 72.08 & 63.56 & 0.19 & 0.76 & 135.01 & 9.249 & 0.821 & -0.0119 & -0.0110 & PI & No\\
                140 & $\,\,0\, \sigma$  & 72.85 & 64.31 & 0.11 & 0.83 & 134.84 & 9.244 & 0.869 & -0.0119 & -0.0109 & PI & No\\
                140 & $+1\, \sigma$ & 73.92 & 65.67 & 0.05 & 0.87 & 134.98 & 9.239 & 0.884 & -0.0115 & -0.0106 & PI & No\\
                140 & $+2\, \sigma$ & 72.85 & 64.47 & 0.01 & 0.86 & 134.85 & 9.243 & 0.884 & -0.0118 & -0.0109 & PI & No\\
                140 & $+3\, \sigma$ & 72.41 & 63.95 & 0.00 & 0.83 & 134.55 & 9.250 & 0.867 & -0.0132 & -0.0122 & PI & No\\
            \hline
                160 & $-3\, \sigma$ & 56.94 & 53.57 & 0.37 & 0.58 & 149.97 & 9.434 & 0.020 & 0.0310 & 0.0308   & CC & Yes\\
                160 & $-2\, \sigma$ & 57.54 & 53.54 & 0.28 & 0.67 & 149.75 & 9.443 & 0.019 & 0.0250 & 0.0248   & CC & Yes\\
                160 & $-1\, \sigma$ & 58.08 & 53.80 & 0.19 & 0.76 & 149.55 & 9.435 & 0.042 & 0.0238 & 0.0237   & CC & Yes\\
                160 & $\,\,0\, \sigma$  & 84.80 & 75.20 & 0.10 & 0.83 & 153.31 & 9.216 & 0.821 & -0.0112 & -0.0103 & PI & No\\
                160 & $+1\, \sigma$ & 84.74 & 75.20 & 0.04 & 0.86 & 153.22 & 9.220 & 0.858 & -0.0124 & -0.0115 & PI & No\\
                160 & $+2\, \sigma$ & 83.70 & 74.31 & 0.01 & 0.85 & 152.53 & 9.219 & 0.853 & -0.0116 & -0.0108 & PI & No\\
                160 & $+3\, \sigma$ & 85.36 & 76.07 & 0.00 & 0.82 & 153.41 & 9.216 & 0.823 & -0.0117 & -0.0107 & PI & No\\
            \hline
        \end{tabular} 
        \footnotesize{
        Column 1: ZAMS mass. Columns from 2 to 9 same as in Table~\ref{tab:HeTable}.
        Column 10: \G1{Core-}. Column 11: \G1{TOT-}. Column 12: final fate of the star, which can be either core collapse (CC) or pair-instability (PI).  Column 13: indicates if stars undergo dredge-up during the evolution.
        }
        \label{tab:HTable3} 
    \end{table*}

\section{Pure-He {\sc parsec}--{\sc mesa} tracks comparison}
\label{appendix:Mesa}
    
	\begin{figure*}
    	\includegraphics[width=\textwidth]{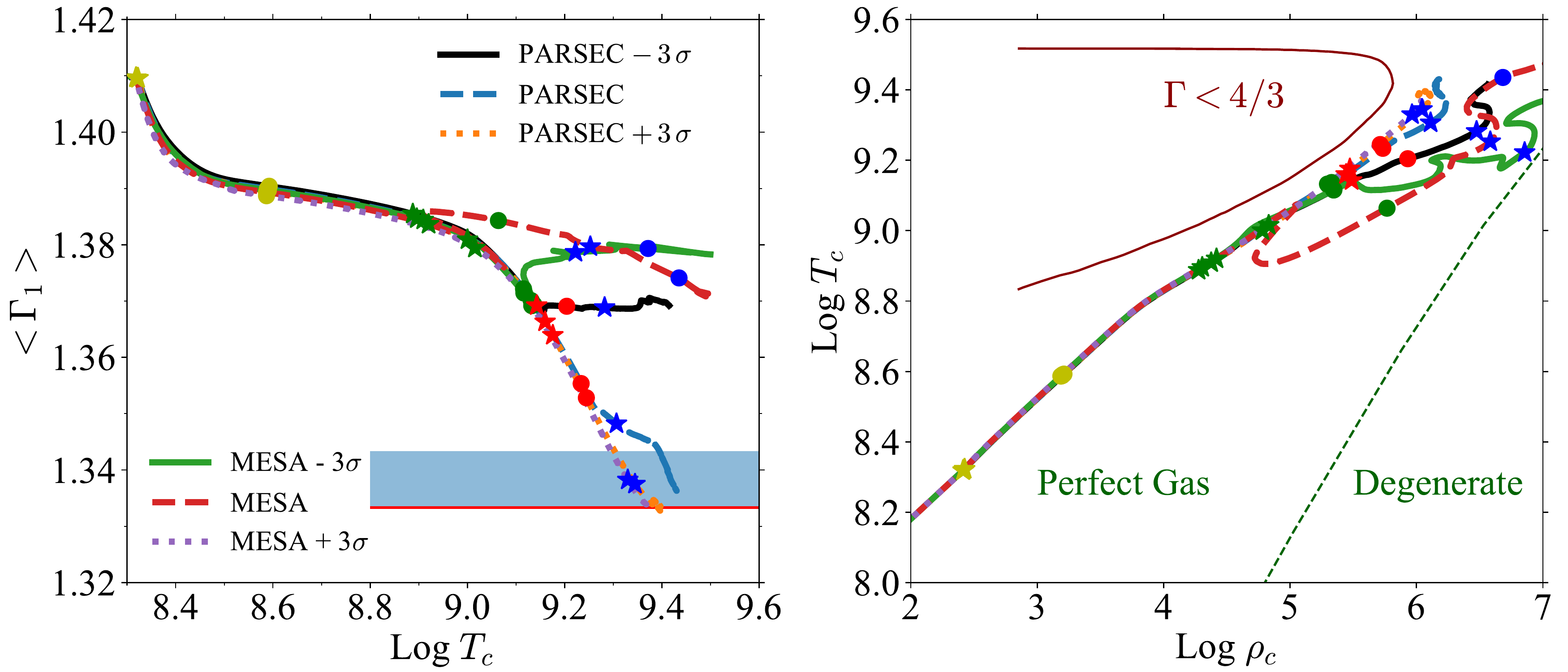}
        \caption{
        	Left-hand panel: evolution of the weighted first adiabatic exponent, \G1, of M$_{\rm He-ZAMS}$ = 40~\Msun\ pure-He stars computed with {\sc mesa} and {\sc parsec} evolutionary codes. {\sc mesa} tracks are from sets computed by \citet{Farmer2020}. Solid lines indicate tracks computed with $-3\, \sigma$ rates, dashed lines show tracks with $0 \, \sigma$, while dotted lines indicate tracks with $+ 3 \sigma$. The red line indicates \G1 = 4/3. The blue region indicate the zone between 4/3 and 4/3 + 0.01, zone in which we consider our tracks unstable. 
        	Right-hand panel: $\rho_{\rm c}$-T$_{\rm c}$ diagram of same tracks of the left-hand panel. In both the left-hand and right-hand panels we use the same symbols as in Fig.~\ref{fig:PureHe_gamma1} and~\ref{fig:H100_gamma1}.
                }
        \label{fig:Comp_Mesa_Parsec}
	\end{figure*}
	Fig.~\ref{fig:Comp_Mesa_Parsec} shows the comparison between 40~\Msun\ pure-He models with varying \CagO\ rates computed in this work and pure-He models of 40~\Msun\ computed by \citet{Farmer2020} with {\sc mesa} stellar evolutionary code. The {\sc mesa} tracks evolve until the final core collapse, but in the figures we show only  tracks until Log (T$_{\rm c}$/K) = 9.5 or \G1\ = 4/3.
    In spite of a different assumed treatment by the two codes for convection, winds, opacity and other physical processes, the evolution during the CHeB phase of the two sets of models is very similar (as shown in \G1\ -- T$_\mathrm{c}$ and $\rho_\mathrm{c}$ -- T$_\mathrm{c}$ diagrams). 
    {\sc parsec} and {\sc mesa} models computed with +3 $\sigma$ show a very similar evolution during all the burning phases and they both become unstable shortly after oxygen ignition. When \G1\ reaches 4/3, {\sc parsec} and {\sc mesa} models have a M$_{\rm Pre}$ of 38.7~\Msun\ and 39.8~\Msun, respectively. The {\sc mesa} model follows the evolution through the PPI, in which the star loses mass and then returns dynamically stable. After the pulsating phase, the star undergoes the final core collapse leaving a BH of 34.8~\Msun. 
        
    {\sc parsec} and {\sc mesa} models with 0 $\sigma$ and -3 $\sigma$ start to behave differently after the ignition of carbon in the core.
    During the core carbon burning phase, the {\sc mesa} model with 0$\sigma$ develops an interaction between the external part of the carbon core and the bottom of the He envelope, that extracts C from the core. This core-envelope interaction stabilizes the model that avoids the PPI and evolves directly to the CC, leaving a BH of 39.8~\Msun\  (Table~\ref{tab:MesaTable}).
    The correspondent {\sc parsec} model, does not undergo the core-envelope interaction and the model become unstable after the ignition of oxygen in the core. This model enters the PPI regime.
    After the depletion of carbon in the core, both {\sc parsec} and {\sc mesa} models with $-3$ $\sigma$ start to burn carbon in a shell above the stellar centre. As already discussed in Section~\ref{sec:Tracks}, the C burning shell sustains the stellar envelope and prevents the collapse of the star, allowing the core oxygen burning in a non-explosive way. At the end of the oxygen burning the {\sc parsec} model has M$_{\rm Pre} =$ 38.9~\Msun\ and can evolve until the final CC. The {\sc mesa} model follows the evolution through the core collapse and the final black hole mass is M$_{\rm BH} = 39.8$~\Msun. Table~\ref{tab:MesaTable} lists some properties of the 40~\Msun pure-He {\sc mesa} models.

    \begin{table} 
        \caption{Pure-He models' with 40~\Msun\ from \citep{Farmer2020}. See text for details.}
        \begin{center}
         
        \begin{tabular}{lcccccc} 
            \hline\hline
            Rate & M$_{\rm He}$ & M$_{\rm CO}$ & XC$_{c-He}$ & XO$_{c-He}$ & M$_{\rm BH}$ & Fate \\

              & [\Msun] & [\Msun] &  &  & [\Msun] & \\
            \hline
            $-3\, \sigma$ & 39.87 & 35.39 & 0.37 & 0.59 & 39.86 & CC\\
            $\,\,0\, \sigma$ & 39.86 & 35.26 & 0.12 & 0.84 & 39.85 & CC\\
            $+3\, \sigma$ & 39.84 & 35.07 & <0.01 & 0.86 & 34.85 & PI\\
            \hline
        \end{tabular}    
        \end{center}
        \footnotesize{The first column reports the rate assumed for the \CagO\ reaction, the second and third ones report the masses of the helium core and CO core at the end of the He-MS, respectively. The fourth and fifth columns list the carbon and oxygen central mass fractions at the end of the He-MS. The sixth column shows the final black hole mass (M$_{\rm BH}$), and the last column reports the final fate of the star: core collapse (CC) or pair instability (PI)}.
        \label{tab:MesaTable} 
    \end{table}

    %


\bsp	
\label{lastpage}
\end{document}

%% file: newcommands.tex
\newcommand{\Hp}{\mbox{H$_\mathrm{P}$}}

\newcommand{\Msun}{\mbox{M$_{\odot}$}}

\newcommand{\lov}{\mbox{$\lambda_\mathrm{ov}$}}
\newcommand{\amlt}{\mbox{$\alpha_\mathrm{MLT}$}}

\newcommand{\beq}{\begin{equation}}
\newcommand{\eeq}{\end{equation}}
\newcommand{\beqa}{\begin{eqnarray}}
\newcommand{\eeqa}{\end{eqnarray}}

\newcommand{\CagO}{$^{12}$C($\alpha$, $\gamma$)$^{16}$O}
\newcommand{\COf}{$^{12}$C/$^{16}$O}
\newcommand{\G}[2]{\mbox{$\langle \Gamma_{#1} \rangle_{\rm #2}$}}
\newcommand{\fco}{\mbox{${\rm f}_{\rm CO}$}}